\documentclass[prd,aps,preprint,amsmath,nofootinbib,amssymb,eqsecnum,showkeys,tightenlines]{revtex4-1}
\usepackage{slashed}
\usepackage{epsfig,latexsym,cancel,amssymb,amsmath,verbatim,mathrsfs}
\usepackage{color}
\usepackage{graphicx}

\def\ra{\rightarrow}
\def\L{\left(}
\def\R{\right)}
\def\wt{\widetilde}
\def\ld{\lambda} 
\def\f{\frac}
\newcommand{\GeV}{\text{GeV}}
\newcommand{\eff}{\text{eff}}
\newcommand{\col}{\text{col}}
\newcommand{\sw}{\text{sw}}
\newcommand{\turb}{\text{turb}}

\begin{document}


\preprint{KIAS-P17047}

\title{
Strong First Order EWPT \& Strong Gravitational Waves \\ in $Z_3$-symmetric Singlet Scalar Extension
}

\author{Zhaofeng Kang}
\email[E-mail: ]{zhaofengkang@gmail.com}
\affiliation{School of Physics, Korea Institute for Advanced Study,
Seoul 130-722, Korea}
\affiliation{School of physics, Huazhong University of Science and Technology, Wuhan 430074, China}

\author{P. Ko}
\email[E-mail: ]{pko@kias.re.kr}
\affiliation{School of Physics, Korea Institute for Advanced Study,
Seoul 130-722, Korea}

\author{Toshinori~Matsui} 
\email[E-mail: ]{matsui@kias.re.kr}
\affiliation{School of Physics, Korea Institute for Advanced Study,
Seoul 130-722, Korea}

\date{\today}

\begin{abstract} 
The nature of electroweak (EW) phase transition (PT) is of great importance. It may give a clue to 
the origin of baryon asymmetry if EWPT is strong first order. Although it is second order within the standard model (SM), a great many extensions of the SM are capable of altering the nature. 
Thus, gravitational wave (GW), which is supposed to be relics of strong first order PT, is a good 
complementary probe to new physics beyond SM (BSM). We in this paper elaborate the patterns 
of strong first order EWPT in the next to simplest extension to the SM Higgs sector, by introducing 
a $Z_3$-symmetric singlet scalar. We find that, in the $Z_3$-symmetric limit, the tree level barrier 
could lead to strong first order EWPT either via three or two-step PT.  Moreover, they could produce 
two sources of GW, despite of the undetectability from the first-step strong first order PT for the near future GW experiments. But the other source with significant supercooling which then gives rise to 
$\alpha\sim{\cal O}(0.1)$ almost can be wholly covered by future space-based GW interferometers such as eLISA, DECIGO and BBO.  
\end{abstract}


\maketitle

\section{introduction}

The early universe must have experienced electroweak phase transition (EWPT), although we do 
not know its nature, namely whether it is first or second order phase transition. Within the standard 
model (SM) of particle  physics, it must proceed continuously owing to the heaviness of the 
SM(-like) Higgs boson. However, the SM is commonly believed to be just a low energy effective 
theory of more fundamental theory, and in the more complete theory EWPT may be first order. 
Actually, we do have a strong motivation that this  may be the case in the context of EW 
baryogenesis (EWBG)~\cite{EWBG,Morrissey:2012db}, which relies on the departure of thermal 
equilibrium furnished by the first order EWPT.  We should further require EWPT to be strong 
enough  so that the baryon asymmetry would not be washed out. This would amount to imposing 
the following condition: 
\begin{eqnarray}\label{eq:SFOEWPT}
\langle h\rangle_*/T_*\gtrsim 1,
\end{eqnarray}
with $T_*$ being the temperature of EWPT and $\langle h\rangle_*$ the vacuum expected value (VEV) of the SM 
Higgs field $h$ at $T_*$.  Therefore, SM extensions that could realize strong first order EWPT (SFOEWPT) is of great interest.  These extensions could help to build a barrier between the EW vacuum and a metastable vacuum at tree or loop level~\cite{Morrissey:2012db,Chung:2012vg}. In order to generate a thermal cubic term for $h$, the latter mechanism usually needs many bosonic degrees of freedom that couple to the Higgs doublet with strength $\gtrsim {\cal O}(1)$, which thus tends to violate perturbativity near the weak scale. In addition, it suffers from gauge dependence issue~\cite{Patel:2011th}. On the other hand, the former mechanism utilizing a tree level barrier can avoid these drawbacks. This mechanism is most easily implemented in the extended Higgs sectors by a (supersymmetric) singlet $S$, containing effective tree-level cubic terms $\sim S^3+S|H|^2$ with $H$ the SM Higgs doublet~\cite{NMSSM:PT1,NMSSM:PT2,NMSSM:PT3,NMSSM:PT4,Profumo:2007wc,Noble:2007kk,Ashoorioon:2009nf,2step1,NMSSM:PT5,Huang:2014ifa,Fuyuto:2014yia,Profumo:2014opa,Chen:2014ask,Kanemura:2015fra,Tenkanen:2016idg,Kanemura:2016lkz,Huang:2016cjm,Hashino:2016xoj,Bian:2017wfv}. 

If the extended Higgs sector respects some symmetry such as $Z_2$, under which $S\ra -S$ and $H\ra H$, 
an alternative way to the desired tree level barrier  is available in the symmetric limit where $S$ does not acquire 
VEV at the present universe~\cite{2step1,2step2,Curtin:2014jma,Vaskonen:2016yiu,Chala:2016ykx,Chao:2017vrq,Beniwal:2017eik}.  Such a scenario is associated 
with multi-step PT's~\cite{Land:1992sm,multiPT,Hammerschmitt:1994fn,Huang:2014ifa,Chao:2017oux} which may happen if there is (was) metastable vacua (denoted as 
$\Omega_{\rm meta}$ which breaks some symmetry like $Z_2$ in this example) except for the desired one ($\Omega_{\rm EW}$) showing  EWSB: The universe may have been once in the intermediate phase $\Omega_{\rm meta}$ and then tunneled through a tree level barrier to the phase $\Omega_{\rm EW}$, recovering the $Z_2$ symmetry.~\footnote{It is tempting to regard $S$ as the dark matter (DM) candidate. However, it fails, at least being the dominant DM component because the relic density is suppressed owing to the large singlet-Higgs coupling which is required by SFOEWPT~\cite{2step2,Curtin:2014jma}.} Since the barrier is present at tree level, SFOEWPT  could be realized simply by reducing the vacuum energy gap between $\Omega_{\rm meta}$ and $\Omega_{\rm EW}$, which would lower the critical temperature~\cite{Huang:2014ifa,Curtin:2014jma}. However, very recently Ref.~\cite{Kurup:2017dzf} raised the question on easy SFOEWPT in that model. 
Previous studies did not consider if there is a bounce solution giving $S_3(T)$ which satisfies 
$S_3(T_*)/T_*\sim {\cal O}(140)$, the condition for bubble nucleation.   As noticed in Ref.~\cite{Kurup:2017dzf}, 
this condition could rule out a large portion of the parameter space, in particular the most attractive one with weak 
couplings that leads to the so called nightmare scenario at colliders. As a matter of fact, SFOEWPT can be probed 
via the Higgs self-coupling measurement~\cite{Kanemura:2004ch,Noble:2007kk,AKS,Kanemura:2011fy,Tamarit:2014dua,Kanemura:2014cka,Hashino:2015nxa,Kakizaki:2015wua,Hashino:2016rvx,Huang:2016cjm,Hashino:2016xoj,Kobakhidze:2015xlz,Lewis:2017dme,Chen:2017qcz}, however, a sufficiently good precision is unlikely until the next generation of colliders such as the International Linear Collider (ILC)~\cite{ILC} and the Compact Linear Collider (CLIC)~\cite{CLIC}. The plan of ILC shows that the Higgs self-coupling can be determined with $10\%$ accuracy by upgrading the center-of-mass energy to $\sqrt{s}=1~{\rm TeV}$~\cite{ILCHiggsWhitePaper,Moortgat-Picka:2015yla,Fujii:2015jha}. 

Maybe the only available test to the nightmare scenario is the gravitational wave (GW) signal. 
It is well known that GWs were emitted after the bubble collision during first order PT (FOPT)~\cite{GWs:early1,GWs:early2,GWs:early3}. The resulting GW spectrum shows a characteristic peak which is related to the PT temperature $T_*$, or  the bubble nucleation temperature. Therefore, EWPT, which would have taken place at $T_*\sim 100$ GeV with  the corresponding peak frequency $\sim 10^{-5}$ Hz, may leave imprints at the GW observation experiments~\cite{Caprini:2015zlo,Cai:2017cbj}. On the other hand, encouraged by the discovery~\cite{aLIGO} of  GWs by LIGO~\cite{Harry:2010zz}, the next generation detectors: eLISA~\cite{Seoane:2013qna}, DECIGO~\cite{Kawamura:2011zz} and BBO~\cite{Corbin:2005ny}, designed to be sensitive to GW density $\Omega_{\rm GW}  h^2\gtrsim 10^{-16}-10^{-10}$ (depending on frequency $\simeq 10^{-3}-10^{-1}$ Hz), will be launched in the near future~\cite{Caprini:2015zlo}. Thus, SFOEWPT can be examined by the future GW interferometers. Once GW signals with that frequency are observed, it is about to yield deep implications to new physics. 

In this paper we study an obvious alternative to $Z_2$, the $Z_3$-symmetric singlet scalar extension. 
Theoretically, $Z_3$ is not inferior to $Z_2$ at all, both of them frequently used in new physics model 
building~\cite{Ma:2007gq,Kang:2010ha,Belanger:2012zr,Ko:2014nha,Aoki:2014cja,Guo:2015lxa,Cai:2016hne,Arcadi:2017vis,Baker:2016xzo}. Although the next simplest extension to the SM Higgs sector, it is strange that such a simple model has not been studied yet in the context of EWPT and GW. Besides, as one of our original but abortive idea, the $Z_3$ extension is supposed to have one more GW source since the first-step PT is supposed to be first order due to the appearance of cubic term $S^3$, and therefore it is a good case in point to demonstrate novel twin-peak GW. Unfortunately, this source is beyond the sensitivity of coming experiments. Still, after employing analytical and numerical methods, we find that the model shows remarkable new features such as the three-step PT, realizing SFOEWPT with strong GW signals which can be well  received by eLISA and DECIGO.

The paper is organized as follows: In Section II we introduce the model; In Section III we analyze 
the vacuum structure at zero temperature as well as its evolution. In Section IV we study the 
gravitation waves from multi-step PT's in our $Z_3$ extension of the SM.   Then conclusions and 
discussions are put in the final section.

\section{$Z_3$ symmetric Higgs sector with a singlet scalar} 
 
Despite of receiving much more attention, a $Z_2$ discrete symmetry does not take obvious advantage 
over $Z_3$ from theoretical viewpoints. On the contrary, we will see that the Higgs sector extended by a 
$Z_3$-symmetric scalar has more interesting features. Besides, this model is one of the very few model 
which can be studied analytically, at least in the leading order. 
 
On top of the Higgs doublet, the Higgs sector contains an isospin complex singlet scalar $S$ transforming 
as $S\ra e^{i2w} S$ with $w=\pi/3$ under $Z_3$, while the SM fields including the SM Higgs doublet $H$ 
are neutral under $Z_3$. Then the most general renormalizable and $Z_3$-symmetric scalar  potential
$V(H,S)$ is given by
\begin{eqnarray} 
V(H,S)&=&-\mu_h^2|H|^2-\mu_s^2|S|^2+{\ld_h}|H|^4+ \ld_s|S|^4\cr
&+&\sqrt{2}\L \f{A_s}{3} S^3+h.c.\R+\ld_{sh}|H|^2|S|^2 .
\end{eqnarray}
Compared to the $Z_2$-symmetric model, there is just one more parameter describing the cubic term $A_sS^3$, and it will give rise to distinguishable difference from the $Z_2$-symmetric model. In this paper we do not consider the possibility that $S$ makes the DM candidate~\cite{Belanger:2012zr}, because we failed in finding viable parameter space with $\ld_{sh}\sim {\cal O}(0.01)$ that is necessary to accommodate correct DM phenomenology. But we would like to point out that this Higgs sector could be a part of more complete model where DM is provided by another ingredients of the complete model. For example, one can consider two loop radiative neutrino mass models with DM running in the loop~\cite{Ma:2007gq,Aoki:2014cja,Ding:2016wbd}, where DM candidate is well furnished, for instance, a Dirac fermion or even another singlet scalar lighter than $S$. Since this DM phenomenology involves other parameters and can be irrelevant to PT, we will not enter its details in this paper, keeping in mind that the $Z_3$-symmetric extension could be a good simplified model for many BSM for DM and neutrino physics.

Expanding the scalar fields around their classical backgrounds, e.g. $S=(s_{\rm cl}+h_s+ia_s)/\sqrt{2}$ 
and so on, one gets the tree level Higgs potential after dropping the subscripts ``{\rm cl}":
\begin{eqnarray}\label{eq:tree}
V_0(h,s)&=&-\f{\mu_h^2}{2}h^2+\f{\ld_h}{4} h^4+\f{\ld_{sh}}{4}h^2s^2
-\f{\mu_s^2}{2}s^2+\f{A_s}{3} s^3+\f{ \ld_s}{4}s^4.
\end{eqnarray}
The vacuum stability condition reads as
\begin{eqnarray} 
\ld_s>0,~~  \ld_h>0~~ {\rm and}~~\ld_{sh}>-2\sqrt{\ld_s\ld_h}.
\end{eqnarray}
At zero temperature $T=0$, the SM Higgs parameters  are fixed to be $\ld_h\approx m_h^2/(2v^2)=0.13$ 
up to radiative corrections,  with $v=246~\GeV$ and $\mu_h^2\approx (88.4~\GeV)^2$ for 
$m_h=125$ GeV. For later use, we give the explicit forms for the field dependent mass squared matrix 
for the CP-even/odd and three Goldstone bosons:
\begin{eqnarray} 
M^2 (h, s, T)&=&\left(\begin{array}{cc}3\ld_hh^2+\f{\ld_{sh}}{2}s^2-\mu_h^2 & \ld_{sh}sh \\ \ld_{sh}sh & 2A_s s+3\ld_ss^2+\f{\ld_{sh}}{2}h^2-\mu_s^2\end{array}\right) \cr
&& +
 \frac{T^2}{48}
 \left(
\begin{array}{cc}
9g^2+3g'^2+12(y_t^2+y_b^2)+24\ld_h+4\ld_{sh}&0 \\
0&16\ld_s+8\ld_{sh}
  \end{array}
 \right)
,\cr
M_{a_s}^2 (h, s, T)&=&-\mu_s^2-2A_ss+\ld_ss^2+\f{\ld_{sh}}{2}h^2
+\frac{T^2}{48}(16\ld_s+8\ld_{sh}), \cr
M_{G}^2 (h, s, T)&=&-\mu_h^2+\ld_hh^2+\f{\ld_{sh}}{2}s^2
+\frac{T^2}{48}(9g^2+3g'^2+12(y_t^2+y_b^2)+24\ld_h+4\ld_{sh}),
\end{eqnarray}
where for completeness we also incorporate the leading order thermal masses. In addition, the field dependent masses for the weak gauge bosons and top quarks are given by, for example, Ref.~\cite{Hashino:2016rvx}. These equations shall furnish the starting point of the loop corrections employed in the program {\tt CosmoTransitions}~\cite{Wainwright:2011kj}, which we adopt to study the phase transitions in the model described in this section.

\section{Multi-step phase transitions and SFOEWPT}

Before the start of vacua structure analysis, it is useful to brief the concept of vacuum tunneling at finite temperature, which also plays key roles in gravitational wave radiations. Tunneling from a metastable vacuum to the ground state through a barrier follows the picture of bubble expansion. For a given scalar potential $V(\vec\phi, T)$ with $\vec\phi$  denoting a vector of real scalar fields in the multi dimensional fields space, the (critical) bubble can be found by extremizing the Euclidean action $S_E(T)$, which can be done numerically by the program {\tt CosmoTransitions}~\cite{Wainwright:2011kj}. Then, the bubble  nucleation rate per unit volume per unit time will be given by $\Gamma(t)=\Gamma_0(t)\exp[-S_E(t)]$ with the prefactor $\Gamma_0 \sim T^4$~\cite{BubbleN}. One may also write $S_E(T) = S_3(T)/T$ where $S_3 (T)$ is defined as
\[
S_3 (T) \equiv \int d^3x\left[\f{1}{2}(\partial\vec\phi)^2+V(\vec\phi{, T})\right].
\]   
In order for the nucleated vacuum bubbles to percolate through the whole Universe, the nucleation rate per Hubble 
volume per Hubble time should reach the unity
\begin{align}
\frac{\Gamma}{H^4}\Bigg|_{T=T_*} \simeq 1,
\end{align}
which determines the FOPT temperature $T_*$. This condition is converted to the very strict relation $S_3(T_*^{})/T_*^{} =4\ln (T_*/H_*)  \simeq 140-150$. Such a condition may block some FOPT, especially for the case that the barrier is induced by tree level effects. 

\subsection{Vacuum structure: minima in the high temperature expansion}

\subsubsection{Preliminaries} 

\begin{figure}
 \begin{center}
 \includegraphics[scale=0.5]{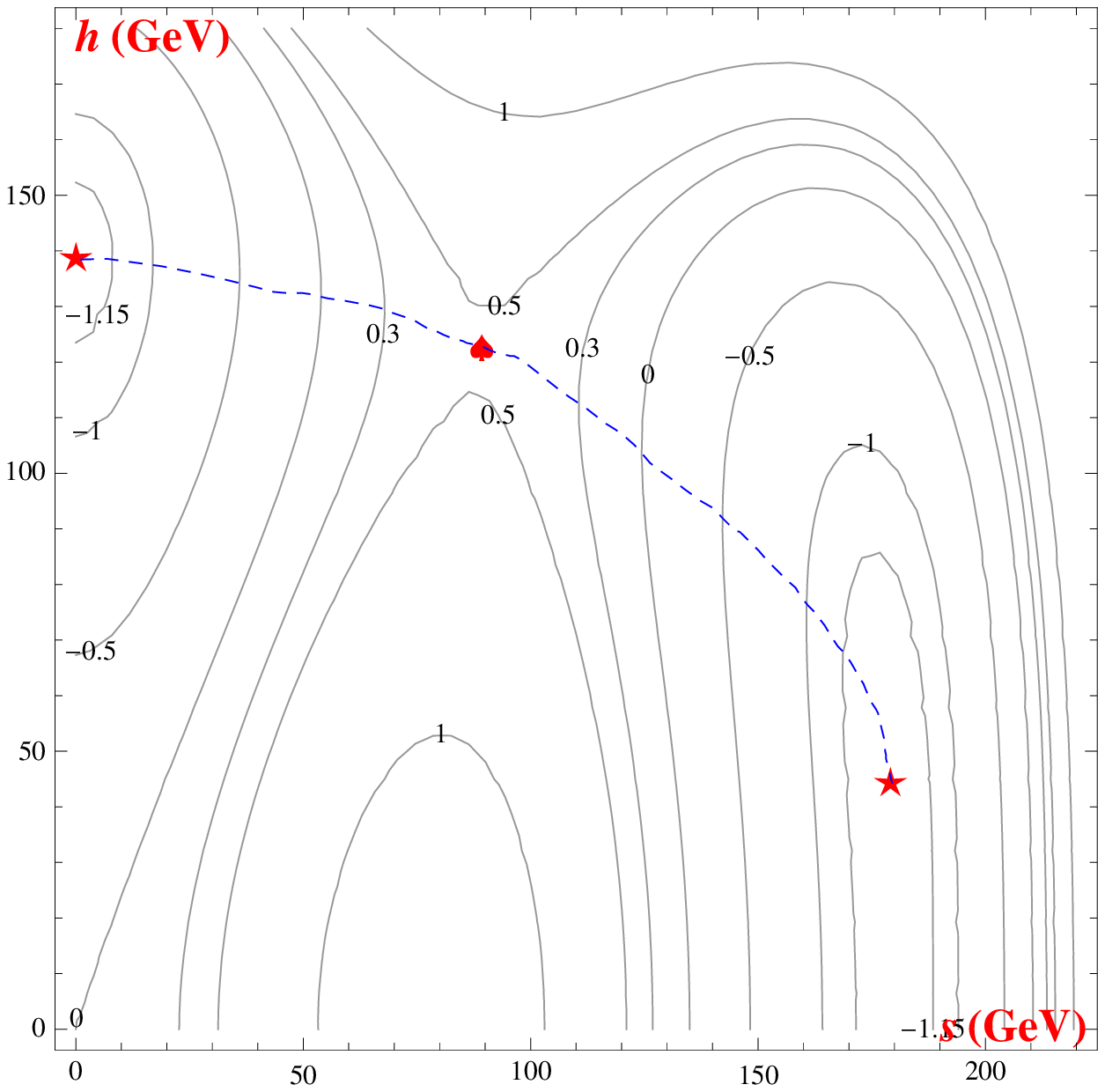}~~~~~~~~~~~~~~
\includegraphics[scale=0.5]{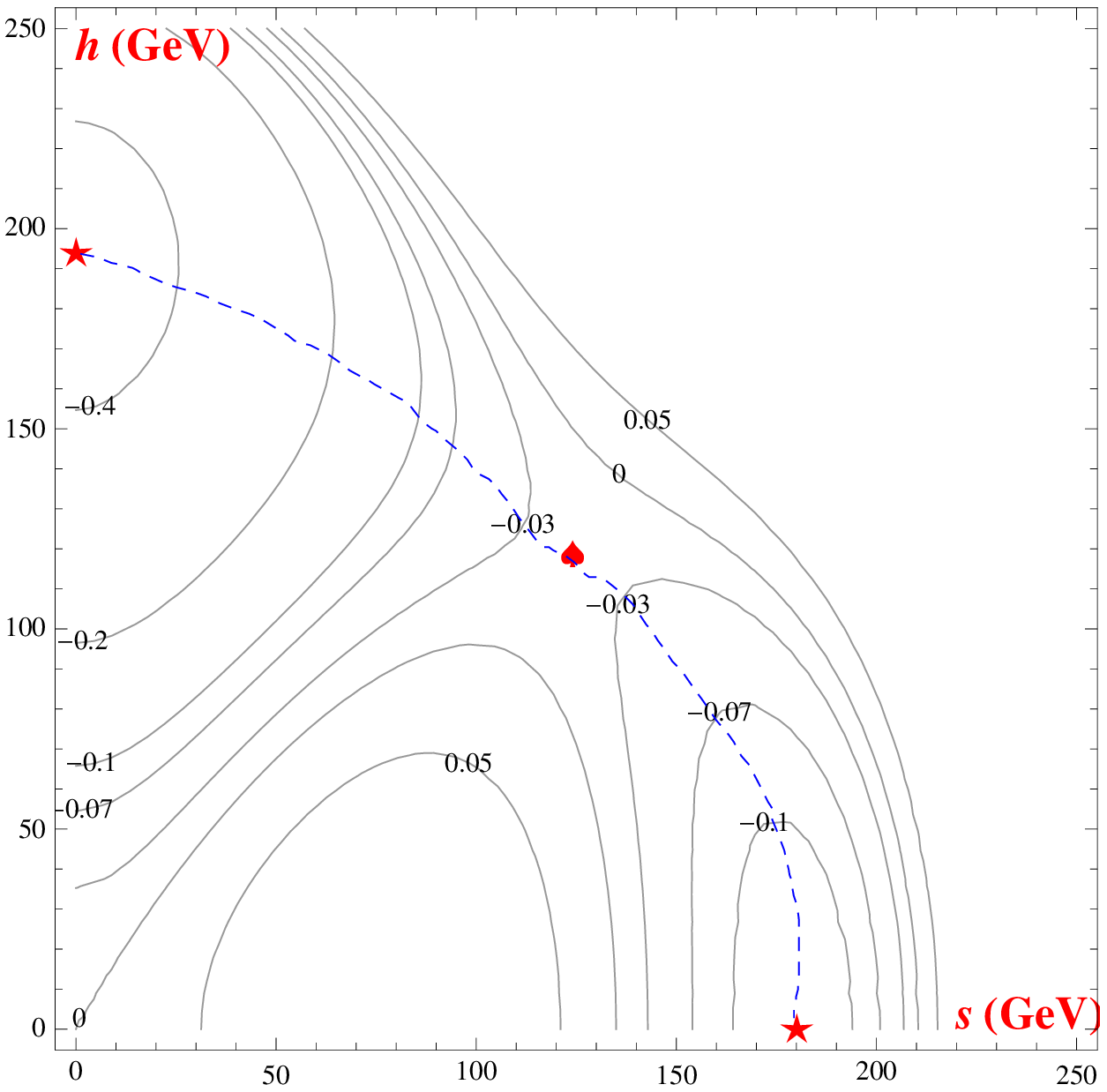}
 \caption{\label{fig:saddle}
Examples for two vacua separated by a tree level barrier: $\Omega_h$ with a metastable $\Omega_{sh}$ 
(left panel) or metastable $\Omega_{s}$ (right panel). The contours are equipotential curves, rescaled by  
large numbers which are of no importance here. The red stars and hearts stand for minima and saddle 
points, respectively. The dashed lines schematically show the tunneling paths from $\Omega_{sh/s}\ra\Omega_h$.}
 \end{center}
\end{figure}

At $T=0$, we are interested in the case where the EWSB but $Z_3$-preserving vacuum $\Omega_h\equiv(\langle h\rangle=v, 0)$ is the ground state, which may be accompanied by a metastable vacuum $\Omega_s\equiv(0, \langle s\rangle\neq0)$ or $\Omega_{sh}\equiv(\langle h\rangle\neq0, \langle s\rangle\neq0)$. The presence of $\Omega_{sh}$ is a new aspect in the $Z_3$-symmetric model compared to the $Z_2$-symmetric model, and it will make possible three-step PT's in our model. In any case, tree level barrier is indeed furnished, as shown in Fig.~\ref{fig:saddle}

To study the vacuum structure and its evolution as $T$ varies, in principle we should find extremes based 
on the effective potential $V_{\eff}(s,h;T)$ at a given $T$, which at one loop level contains several terms 
such as the Coleman-Weinberg (CW) potential and finite temperature corrections~\cite{Morrissey:2012db}. 
But that should render analytical analysis impossible. Therefore, for the purpose of analytic discussion in this Section, we instead work in the high temperature expansion of $V_{\eff}(s,h;T)$, without considering the CW potential \footnote{We shall include all contributions up to one loop for numerical analysis.}. Under this approximation, the tree level effective potential Eq.~(\ref{eq:tree}) turns out to be the most general form, while the finite temperature corrections are simply encoded in the evolutions of the mass squared parameters 
\[
\mu_{s,h}^2(T)=\mu_{s,h}^2-c_{s,h}T^2 ,
\] 
with $\mu_{s,h}^2$ the parameters defined at $T=0$~\footnote{In some context, $\mu_{s,h}^2$ also refers to $\mu_{s,h}^2(T)$ to avoid lengthy expressions, but we believe that the readers can easily understand the exact meaning of $\mu_{s,h}^2$. And the same convention applies to $\eta$ too.} and the coefficients 
$c_s , c_h$ are calculated to be 
\begin{eqnarray} 
c_s&=&\f{1}{12}\L2\ld_{sh}+4\ld_s\R,\\
c_h&=&\f{1}{12}\left(\f{9}{4}g^2+\f{3}{4}g'^2+ 3 y_t^2+6\ld_h+\ld_{sh}\right).
\end{eqnarray}
It is seen that $\mu_s^2(T)$ decreases with $T$ as long as $c_s = 2(\ld_{sh}+2\ld_s )>0$. Note that the linear term $A_s sT^2$ is absent, due to the cancellation between the CP-odd and CP-even contributions, as a result of $Z_3$ symmetry. 
 

\subsubsection{Analysis along the singlet direction}
 
It is illustrative to start the analysis along the singlet direction without taking into account $h$, where the parameter $\eta\equiv A_s^2/(4\ld_s \mu_s^2)$ plays an important role. Under high temperature expansion it evolves as
\begin{eqnarray} 
\eta(T)=\f{\eta}{1-c_s T^2/\mu_s^2}.
\end{eqnarray}
It is straightforward to find out that in different regions of $\eta$, the vacuum structures respectively take the forms of:
\begin{description}
\item[$\eta>0$ ]   There are two minima, separated by the maxima, namely the origin; they are located in the positions 
\begin{eqnarray}\label{eq:v:singlet}
v_{s,\pm}=\f{|A_s|}{2\ld_s}\L -1\pm {\rm sign}(A_s)\sqrt{1+1/\eta}\R.
\end{eqnarray}
For $A_s>0$, $v_-$ is the global minimum, otherwise $v_+$. And we denote the minima as $v_s$. Irrelevant to the sign of $A_s$, the deeper one has negative definite vacuum energy $E_s=\f{-A_s^4}{96\ld^3_s} f_+(\eta)$ with 
\begin{eqnarray} 
 f_+(\eta)=\L\f{\sqrt{\eta}+\sqrt{\eta+1}}{\eta}\R^2
\L \eta+\f{3}{2}+\sqrt{\eta^2+\eta}\R,
\end{eqnarray}
which is a monotonously decreasing function of $|\eta|$ with the lower limit 8. This case does not admit a very small $\ld_{sh}$ since the positive singlet scalar mass requires a large $\ld_{sh}$ like the $Z_2$ case. 

\item[$-1<\eta<0$ ] The two minima vanishes and the origin is the only minimum (or only the Higgs doublet acquiring VEV in the two dimensional field space). Note that at very high $T$ where $\mu^2_s(T)\propto T^2\gg \mu_s^2$ and thus the potential is always in this region. 
\item[$\eta<-1$ ] On top of the origin, there is another minima located at $v_{s,+(-)}$ for $A_{s}<(>)0$ and the maxima between the two minima is $v_{s,-(+)}$. Its vacuum energy is $E_s=+\f{A_s^4}{96\ld^3_s} f_-(\eta)$ with 
\begin{eqnarray}\label{eq:EZ3}
f_-(\eta)=\L\f{\sqrt{|\eta|}+\sqrt{|\eta|-1}}{\eta}\R^2
\L \eta+\f{3}{2}-\sqrt{\eta^2+\eta}\R.
\end{eqnarray}
This function lies above zero for $-\f{9}{8}<\eta<-1$ and crosses zero , the vacuum energy of the symmetric vacuum, at $\eta=-\f{9}{8}$ and stays below zero for even smaller $\eta$. When the two minimums are degenerate, the hight of the barrier separating them is 
\begin{eqnarray}\label{eq:barrier}
\Delta E_b=\f{A_s^4}{96\ld^3_s} f_+(\eta=-9/8)\approx7A_s^4/32\ld_s^3
\end{eqnarray}
moreover, the distance of two vacua is $\propto |A_s|/\ld_s$. 
Therefore, a larger $\ld_s$ and a smaller $|A_s|$ will help to reduce the barrier and facilitate the bubble 
nucleation during the transition from the origin to the $Z_3$ breaking minima. 
\end{description}


\subsubsection{Analysis in the $(s,h)$ space}

\begin{figure}
 \begin{center}
 \includegraphics[scale=0.5]{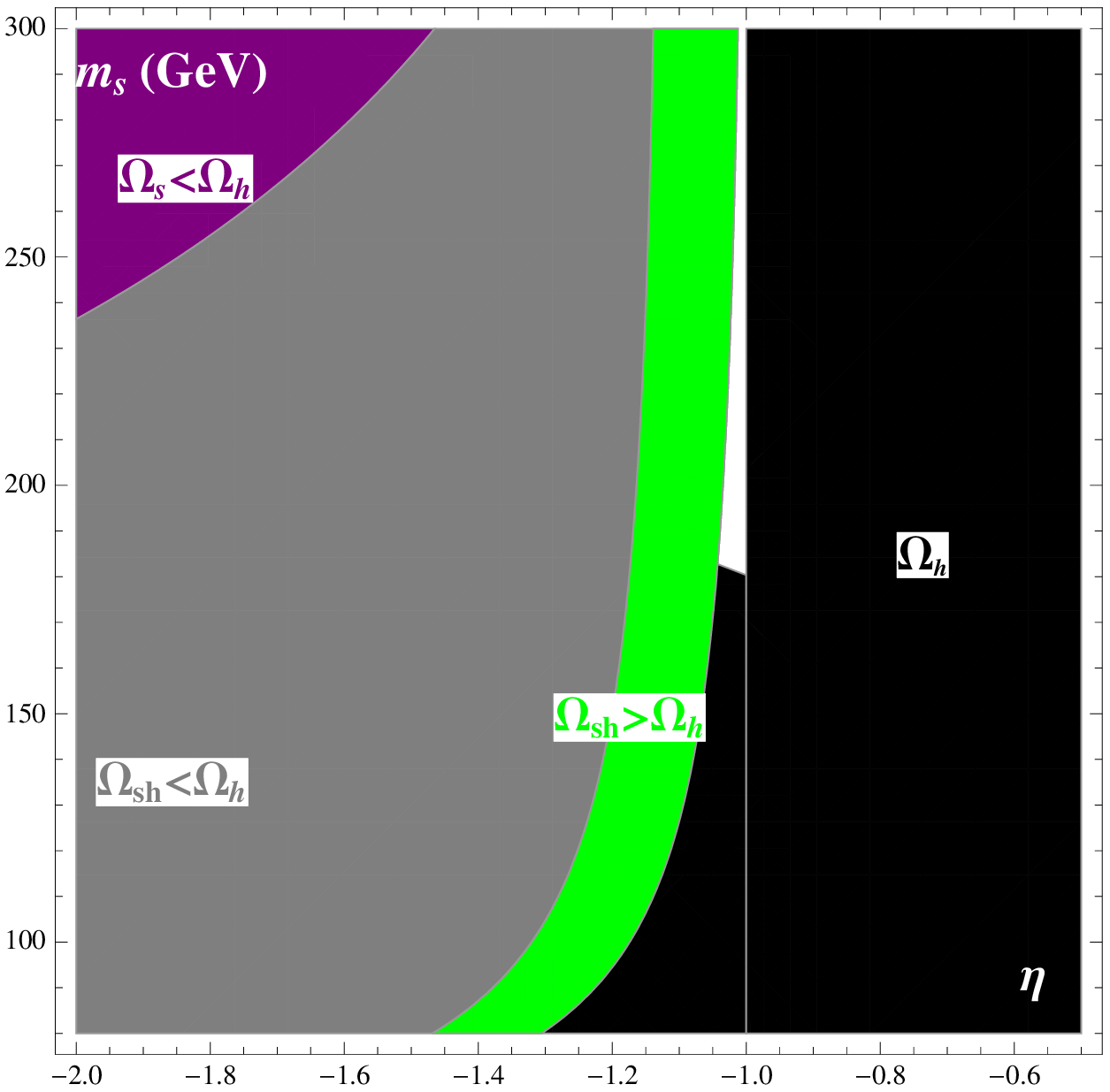}~~~~~~
\includegraphics[scale=0.5]{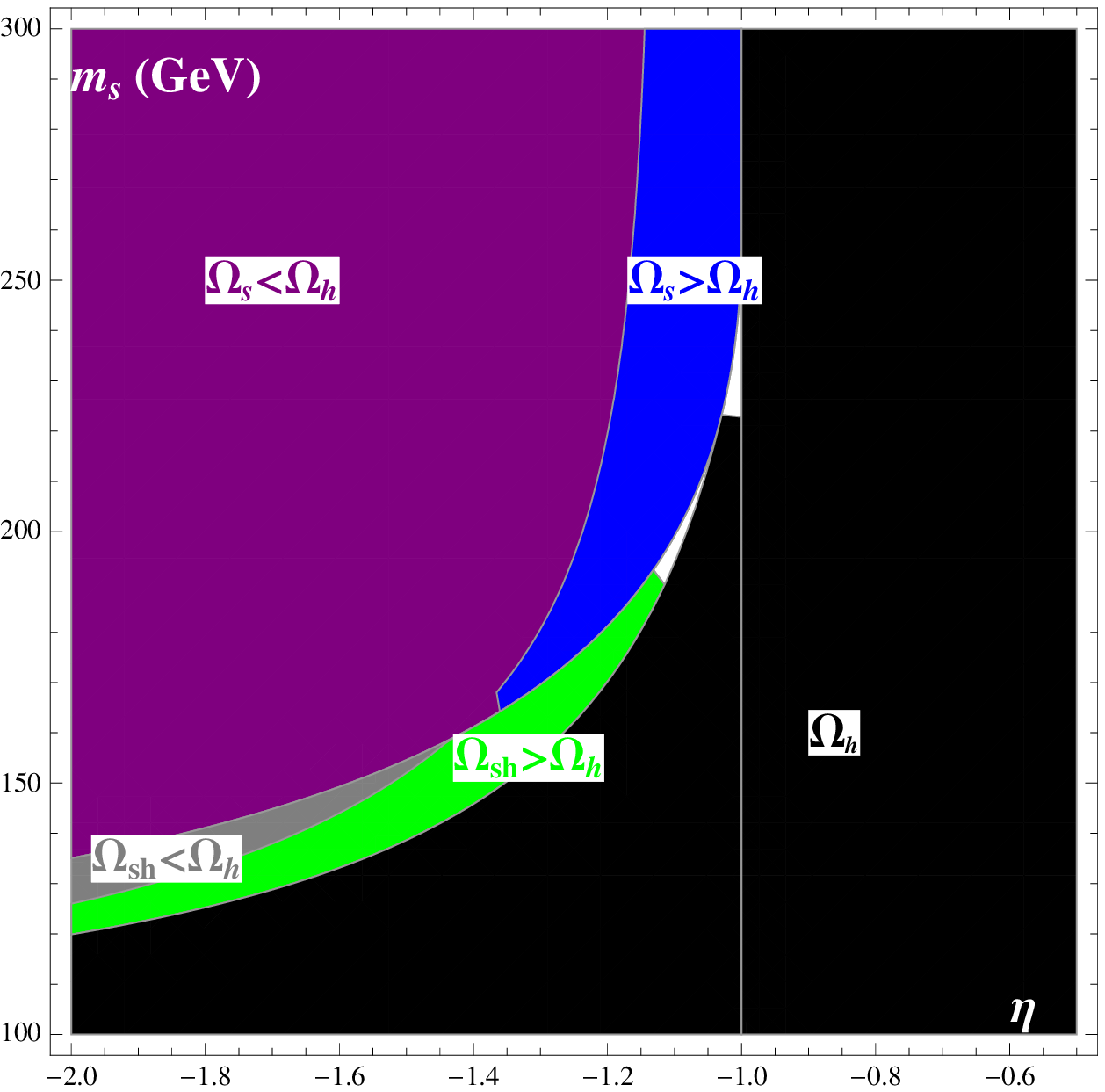}
 \caption{\label{fig:vacu0}
Tree level vacuum structures at zero temperature in the $(\eta, m_s)$ plane for the case $\eta<0$: 
(Left panel) a smaller $\ld_{sh}=0.05$ with $\ld_s=1$; (Right panel)  a larger $\ld_{sh}=0.27$ 
with $\ld_s=1$.  In each colored region, we label (using the same color, a convention applied 
to other figures) all the local minima (with number no lager than two) and their relative orders, says in the 
green regions there are two minima $\Omega_{sh}$ and $\Omega_h$ with the latter the ground state.}
 \end{center}
\end{figure}
\begin{figure}
 \begin{center}
 \includegraphics[scale=0.5]{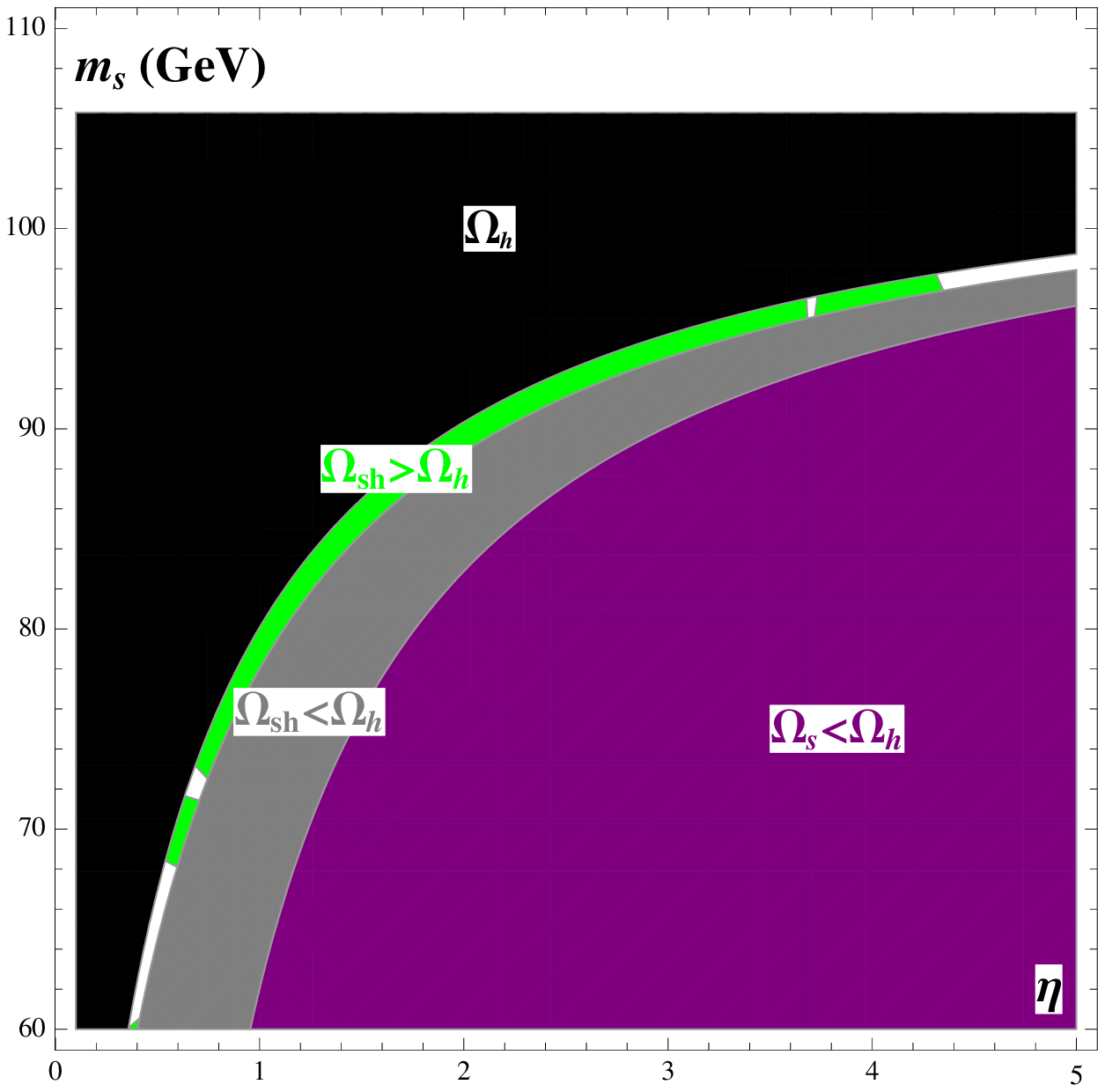}~~~~~~~
\includegraphics[scale=0.5]{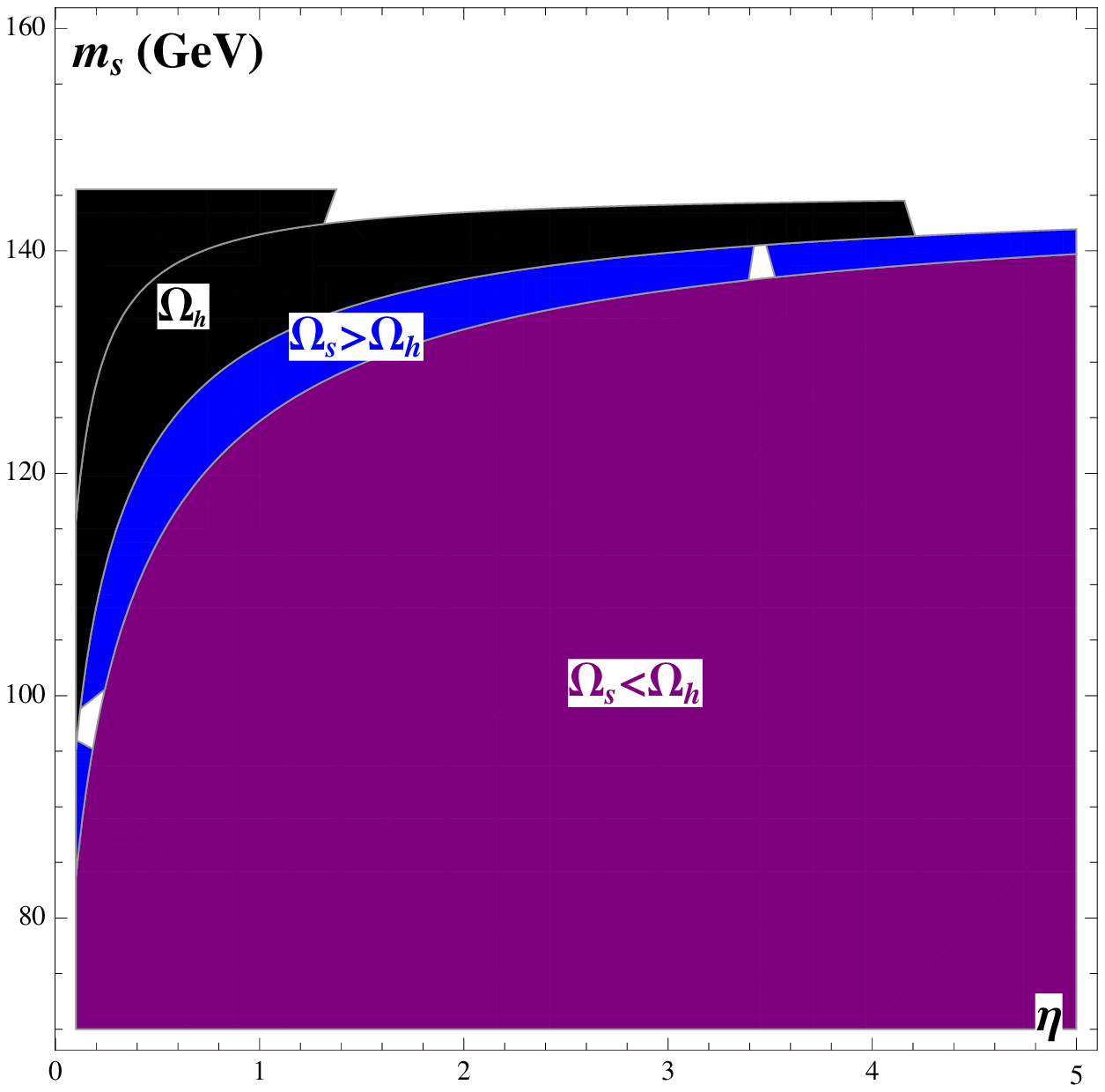}
 \caption{\label{fig:vacu1} Tree level vacuum structures at zero temperature on the $(\eta, m_s)$ plane 
 for the case $\eta>0$:  (Left panel) a smaller $\ld_{sh}=0.37$ with $\ld_s=1$;  (Right panel) a larger 
 $\ld_{sh}=0.7$ with  $\ld_s=1$.}
 \end{center}
\end{figure}

Now let us move to the two-dimensional field space, where new minima, such as the trivial one along the doublet direction $\Omega_h$, arise. We start from the vanishing tadpole equations:
\begin{eqnarray}\label{eq:tadpole}
s\L \ld_s s^2+ A_s s-\mu_s^2+\f{\ld_{sh}}{2}h^2 \R=0,\quad 
h\L \ld_h h^2-\mu_h^2+\f{\ld_{sh}}{2}s^2 \R=0.
\end{eqnarray}
Let us define $\ld_S\equiv 4\ld_h\ld_s-\ld_{sh}^2$, $\mu_S^2\equiv 4\ld_h \mu_s^2-2\ld_{sh}\mu_h^2$, $A_S\equiv 4\ld_h A_s $ and $\eta_S\equiv A_S^2/(4\ld_S\mu_S^2)$. In the limit $\ld_{sh}\ra0$, they are nothing but the rescaling of the corresponding parameters in the singlet sector by a factor $4\ld_{h}$. Then it is straightforward to see that Eq.~(\ref{eq:tadpole}) admits solutions having nonvanishing $\langle s\rangle$ and $\langle h\rangle$ if $1/\eta_S>-1$ and moreover $\ld_{sh}\langle s\rangle^2/2<\mu_h^2$. These solutions are denoted as $\L u_{s,\pm}, v_h\R$ with the singlet VEV similar to Eq.~(\ref{eq:v:singlet}),
\begin{eqnarray} 
 u_{s,\pm}&=&-\f{A_S}{2\ld_S}\L1\pm{\rm sign}(A_S)\sqrt{1+1/\eta_S}\R,\cr
 v_h&=&\pm \sqrt{\L-\ld_{sh}u_{s,\pm}^2+2\mu_h^2\R/2\ld_h},
\end{eqnarray}
with the sign of $v_h$ irrelevant and assigned a positive sign hereafter. The extreme having larger magnitude of singlet VEV has the potential to be the minima $\Omega_{sh}$ while the other one is a saddle point. To realize the potential, one further condition should be fulfilled, i.e., the Higgs doublet becomes tachyonic in $\Omega_s$, otherwise $\Omega_s$ is the minima rather than $\Omega_{sh}$. We will add more details about this point elsewhere. More explicitly, the conditions for accommodating $\Omega_{sh}$ are summarized as 
\begin{eqnarray}\label{eq:con:2}
   \eta\lessgtr\L1-\f{\ld_{sh}^2}{4\ld_s\ld_h}\R\L \f{\ld_{sh}\mu_h^2}{2\ld_h\mu_s^2}-1\R~\&~\ld_{sh}u_s^2/2<\mu_h^2 ~ \&~ \ld_{sh} v_s^2/2<\mu_h^2.
\end{eqnarray}
The first inequality sign takes $``<"$ and $``>"$  for a negative and positive $\mu_s^2$, respectively. Obviously, a smaller $\ld_{sh}$ can readily satisfy all these inequalities and thus $\Omega_{sh}$ is well expected. On the other hand, if $\ld_{sh}$ is not very small (says a few 0.1) and moreover $\ld_s$ is relatively large ($\sim 1$), the appearance of $\Omega_{sh}$ is rare. One can clearly confirm these on the $\eta-m_s$ plane, from the gray and green shaded regions in Fig.~\ref{fig:vacu0} and Fig.~\ref{fig:vacu1}.  


 The potential energy of $\Omega_{sh}$ can be written as the sum of two parts, with one the doublet contribution, 
\begin{eqnarray}\label{eq:Esh}
E_{sh}&=&-\f{A_S^4}{96\ld_h\ld_S^3}f_{sh,+}(\eta_S)-\mu_h^4/4\ld_h,
\end{eqnarray}
while the potential energy of the saddle point has a similar expression with $f_{sh,+}(\eta_S)\ra f_{sh,-}(\eta_S)$ and 
\begin{eqnarray} 
f_{sh,\pm}(\eta_S)&=&2+\f{3}{4\eta_S^2}+\f{3}{\eta_S}\pm2\L1+\f{1}{\eta_S}\R^{3/2}.
 \end{eqnarray}
Like $f_-(\eta)$ introduced before, $f_{sh,+}(\eta_S)$ is a monotonically decreasing function of $\eta_S$ and crosses zero at $\eta_S=-9/8$. The height of the barrier is $\Delta E_{sh}=A_S^4/(24\ld_S^3\ld_h)(1+1/\eta_S)^{3/2}$. Note that the second term in Eq.~(\ref{eq:Esh}), namely $E_h=-\mu_h^4/4\ld_h$, is exactly the vacuum energy of the minima along the doublet direct, $\Omega_h$. Therefore, $\Omega_{sh}$ stays above $\Omega_{h}$ if 
\begin{eqnarray} 
\ld_S= 4\ld_h\ld_s-\ld_{sh}^2<0~ \&~\eta_S<-9/8\cup \eta_S>0~~{\rm or}~~ \ld_S>0~ \&~ -1>\eta_S>-9/8.
\end{eqnarray}
The first can be readily satisfied for a relatively larger $\ld_{sh}$ but smaller $\ld_s$, and it does not impose much severer condition other than Eq.~(\ref{eq:con:2}). However, for a smaller $\ld_{sh}$ and thus $\ld_S>0$, may be just a narrow strip left; see the green shaded regions in the left panels of Fig.~\ref{fig:vacu0} and Fig.~\ref{fig:vacu1}. 

\subsubsection{More on $\Omega_s$ and $\Omega_{sh}$}

It is of importance to address the connection between $\Omega_{sh}$ and $\Omega_s$. Although not proved here \footnote{One may get hints from Ref.~\cite{Espinosa:2010hh} based on a generic real scalar singlet model.}, they do not coexist with each other. Actually, for most parameter profiles, the first and the second conditions in Eq.~(\ref{eq:con:2}) suffice the presence of $\Omega_{sh}$. However, there indeed exists parameter space where those two conditions are satisfied while $\ld_{sh}v_s^2/2>\mu_h^2$ is also met. Then, in this case $\Omega_s$ has the priority and $\Omega_{sh}$ is not present. In the sense of evolution under temperature, ${\cal P}(T)\equiv \ld_{sh}v_s^2(T)/2-\mu_h^2(T)$ somehow can be regarded as an order parameter between the two phases $\Omega_s$ and $\Omega_{sh}$: When ${\cal P}(T)$ crosses zero (from above 0), $\Omega_s$ transits to $\Omega_{sh}$ provided that the other two conditions in Eq.~(\ref{eq:con:2}) are satisfied. In other words, the PT $\Omega_s\ra\Omega_{sh}$ is second order because it is related to continuous variation of ${\cal P}(T)$. However, there is another situation in the region with $\ld_{sh}v_s^2/2-\mu_h^2<0$. That is, maybe those two conditions are violated and then $\Omega_{sh}$ is still not present, only with $\Omega_h$ left. Its implication is that in this case $\Omega_s$ may transit to $\Omega_h$ via second order PT instead of first order. Despite of irrelevant to our study, we did observe such kind of PT in the numerical study.


At $T=0$ the existence of $\Omega_h$ and a metastable $\Omega_{s}/\Omega_{sh}$ does not guarantee SFOEWPT by that tree level barrier. We have to trace the phase evolution up to high temperature. Anyway, these general analysis still guides us to explore the patterns of PT.  

%
\subsection{Vacuum structure: evolution/phase transition~\footnote{
The Coleman-Weinberg and as well as the higher oder terms beyond the high temperature approximation may play non-negligible roles, so that the analysis here is only qualitative, to illustrate the physics points.} 
}


Our universe today is assumed to be in the $\Omega_h$ phase, but it may have experienced other phases at the earlier stage. Due to the multi minima structure in this model, the patterns of PT are rich and complicated. For our purpose, we wish that there was a PT pattern such as two-step PT $\Omega_0\ra \Omega_s(\Omega_{sh})\ra \Omega_h$ or even three-step $\Omega_0\ra \Omega_{s}\ra\Omega_{sh}\ra \Omega_h$. Among them, $\Omega_0$ to $\Omega_{sh}$ being the first-step cannot happen in the parameter space of our interest, and the reason will be clear in the footnote~\ref{fn}. We will find that other cases are feasible in the weak coupling region. But the way to achieve them is quite non-trivial and we may get a clue from tracking the evolutions of $\eta(T)$ and $\mu_h^2(T)$. 

\begin{figure}
 \begin{center}
 \includegraphics[scale=0.5]{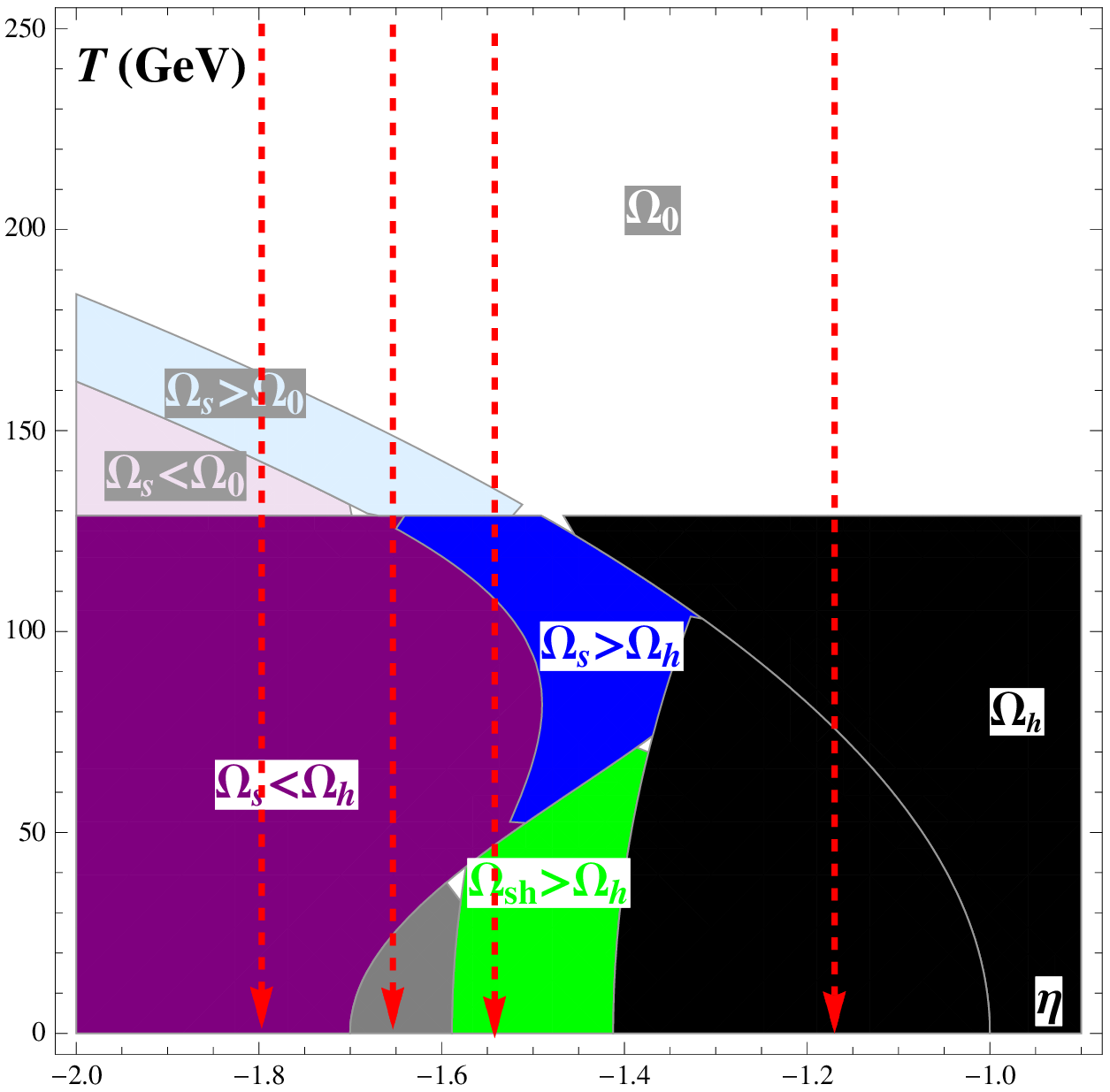}~~~~~~~
\includegraphics[scale=0.5]{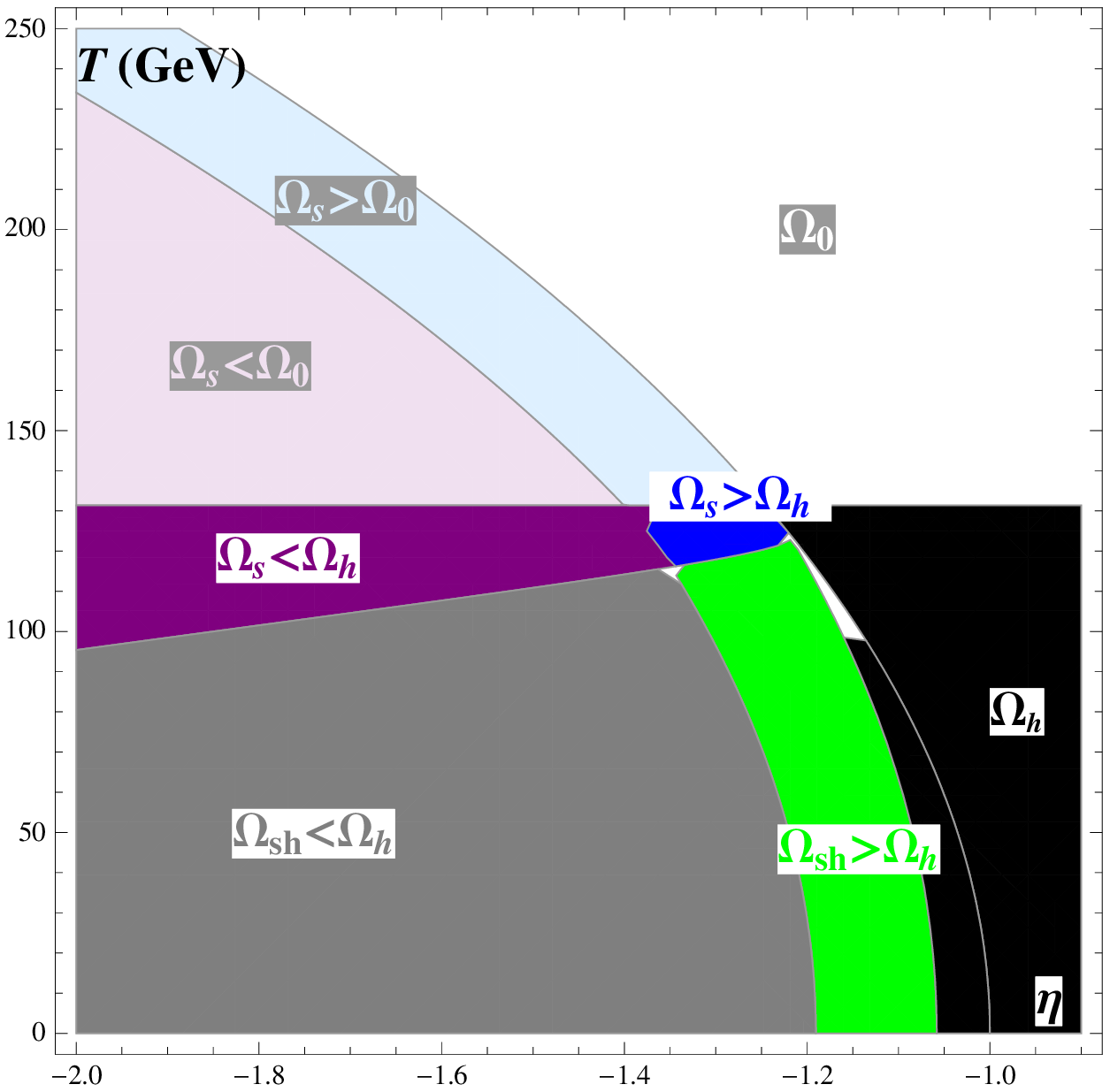}
 \caption{
 Vacuum structure evolution under the high temperature approximation ($\eta<0$ case). (Left panel): 
 $\ld_{sh}=0.27$, $\ld_s=1$ and $m_s=95$ GeV; (Right panel): $\ld_{sh}=0.05$ and $\ld_s=1$ and 
 $m_s=160$ GeV. They are corresponding to Fig.~\ref{fig:vacu0}, but $m_s$ is fixed here now. 
 Dashed arrows indicate the evolutions of minima structure thus possible PT patters, from left to right: 
 1) $\Omega_0\ra  \Omega_s  $; 2) $\Omega_0 \ra \Omega_s\ra \Omega_{sh} $; 3) $\Omega_0 \ra  \Omega_h\ra\Omega_s\ra\Omega_{sh}$; 4) $\Omega_0\ra\Omega_h$. But we keep in mind that higher order terms may change them, in particular the behaviors at the higher temperature.
 } 
 \end{center}
\end{figure}
At very high $T_\infty$, the universe is in the symmetric phase $\Omega_0$ because $\eta(T_\infty)\ra 0^-$ and $\mu_h^2(T_\infty)<0$. When the universe cooled down, local minima $\Omega_h$ or/and $\Omega_s$ appeared when $\mu_h^2(T_h)$ and $\eta(T_s)$ respectively approached 0 and $-1$. Then their birth temperature are estimated to be 
\begin{eqnarray}\label{eq:Tsh}
T_h\approx\f{\mu_h}{\sqrt{c_h}}~~{\rm and} ~~ T_s\approx \f{\sqrt{|\eta+1|}}{\sqrt{c_s}}|\mu_s| ,
\end{eqnarray}
respectively.   The above estimation on $T_h$ is mildy lower than the one calculated in the code, 
which gives $T_h\approx 160$ GeV for a weak coupling $\ld_{sh}<1$. If $T_h>T_s$, in the weak coupling region of $\ld_{sh}\lesssim1$, $\Omega_0$ would immediately roll down to $\Omega_{h}$ at $T_h$. Such a case should be avoided and therefore we should at least impose the condition $T_s>T_h$. In practice, this condition should be strengthened to allow the commencement of PT $\Omega_0\ra \Omega_s$, that is $T_h<T_{s}^*$ with $T_{s}^*$ the critical temperature where $\Omega_s$ became degenerate with $\Omega_s$, which can be determined by 
$\eta(T_s^*)=-9/8$; see the discussion below Eq.~(\ref{eq:EZ3}). However, even given a small enough $\eta$, whether $\Omega_0\ra \Omega_s$ was completed at the PT temperature  $T_s^*\in (T_h, T_s^*)$ depends on if there was a sufficiently large bubble nucleation rate such that $S_3(T_s^*)/T_s^*\lesssim 140$. If not, the universe would be confined in the symmetric phase until the transition to $\Omega_h$. Unfortunately, usually $T_s^*$ is significantly below $T_s^*$ because $\Omega_0\ra \Omega_s$ is hampered by a tree level barrier. Then the strengthened condition $T_s^*>T_h$ is transformed into a bound on the initial $\eta$ (assumed to be negative), 
\begin{eqnarray}\label{eq:con:3}
\eta<\eta_*\L1+\f{c_s}{c_h}\f{\mu_h^2}{|\mu_s^2|}\R,
\end{eqnarray}
where $\eta_*\equiv\eta(T_s^*)<-9/8$ but the concrete value cannot be calculated in this paper.~\footnote{Similarly we can understand why the first-step cannot be $\Omega_0\ra\Omega_{sh}$: The presence of $\Omega_{sh}$ requires $\mu_h(T)^2>0$, which means that $\Omega_{sh}$ is supposed to appear later than $\Omega_{h}$; $\Omega_0$ would immediately roll down to  $\Omega_{h}$ at its presence. A negatively large $\ld_{sh}$ is an exception, but it is a case of no interest.\label{fn}} 
\begin{figure}
 \begin{center}
 \includegraphics[scale=0.5]{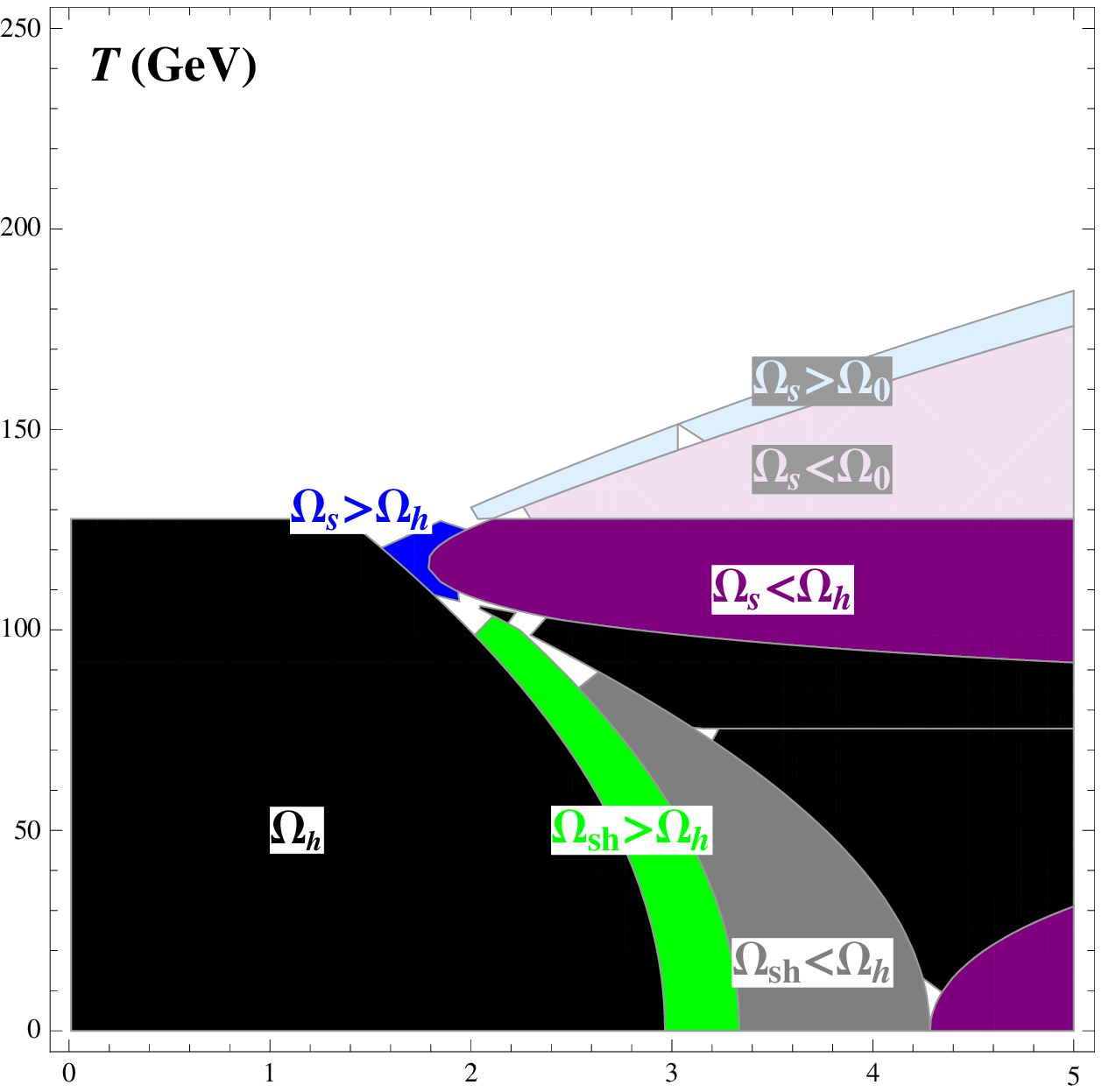}~~~~~~
\includegraphics[scale=0.5]{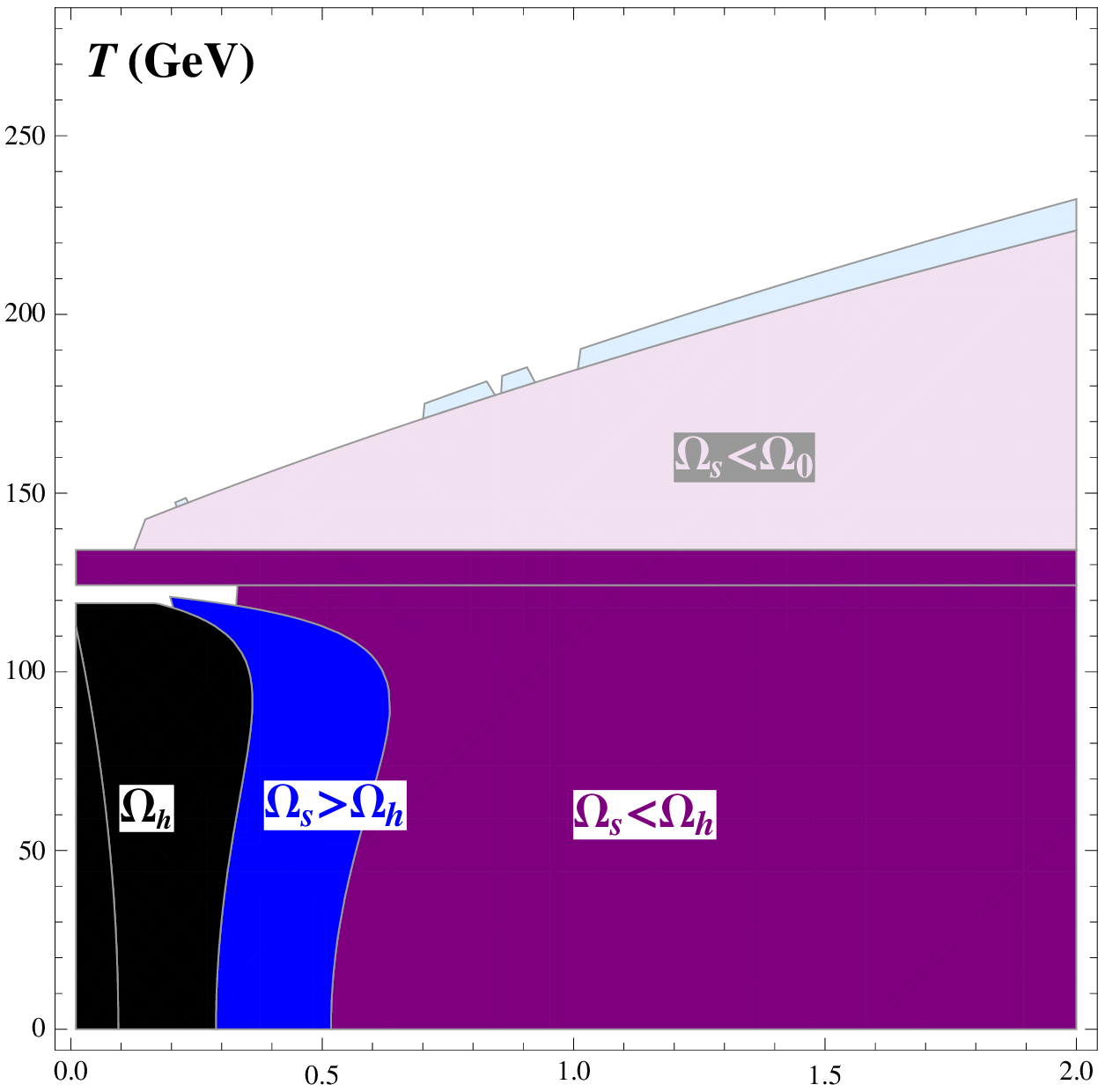}
 \caption{
 Vacuum structure evolution under the high temperature approximation ($\eta>0$ case). 
 (Left panel): $\ld_{sh}=0.37$, $\ld_s=1$ and $m_s=95$ GeV; (Right panel): $\ld_{sh}=0.7$ and 
 $\ld_s=1$ and $m_s=115$ GeV.  } 
 \end{center}
\end{figure}

This raises another possibility which might happen in the $\eta>0$ case. It involves another key temperature $T_s^{0}$ (again assumed to lie above $T_h$) at which $\eta(T_s^{0})\simeq0$ and the origin turned into a maxima; below $T_s^{0}$ the model switched to the $\eta>0$ branch. If $\Omega_0$ was hold above $T_s^0$, then it would transit to $\Omega_s$ via second order PT instead of first order like before. We found that some of the two-step PT samples $\Omega_0\ra\Omega_s\ra \Omega_h$ belongs to this pattern; see the example points labelled as triangle in Fig.~\ref{fig:As-lsh}. 

Even $\Omega_0\ra\Omega_s$ succeeded, it is likely that subsequently $\Omega_s$ second order transited to $\Omega_{sh}$ rather than to $\Omega_h$ directly as explained before. For instance, the vacuum energy of $\Omega_h$, which is $E_h=-\mu_h^4(T)/4\ld_h$, stayed above that of $\Omega_s$, which is $E_s(T)$, or again $S_3(T)/T$ was too large during the window $[T_{sh}, T_s^*]$ with $T_{sh}$ the temperature of appearance of $\Omega_{sh}$; it is also the cross over temperature of transition $\Omega_{s}\ra \Omega_{sh}$. Of importance, to avoid the former, one confronts with an upper bound on $\eta(T)$ derived from $f_-(\eta(T))>-24\ld_s^3\mu^4_h(T)/\ld_hA_s^4\equiv -\epsilon_T$. Despite the lack of an explicit expression, practically, on account of the suppression $\mu_h(T)^4/A_s^4\ll 1$ thus a small $\epsilon_T$ at the higher temperature, it gives 
\begin{eqnarray}\label{eq:con:4}
-\f{9}{8}\L1+0.21 \epsilon_T \R<\eta(T)=\f{\eta}{ 1-c_s T^2/\mu_s^2} <-1.
\end{eqnarray}
Combined with Eq.~(\ref{eq:con:3}) we immediately see that the allowed region for $\eta$ is fairly narrow. In particularly, in the heavy singlet region with $A_s, \mu_s\gg \mu_h$, basically the two-step PT scenario cannot be accommodated in the $\eta<-1$ region. And we did not find successful two-step pattern $\Omega_s\ra \Omega_h$ in this region, at least for weak couplings. 

If $\Omega_s\ra \Omega_h$ fails, we are left with the three-step PT $\Omega_0\ra \Omega_s\ra \Omega_{sh}\ra \Omega_h$. But from the numerical searches it is found that the final-step, in particular in the $\eta<0$ case, tends to suffer a serious bubble nucleation problem and consequently the universe would be confined in the metastable vacuum $\Omega_{sh}$. 
 
\subsection{Numerical samples}

 We use the code {\tt cosmoTransitions}~\cite{Wainwright:2011kj} for numerical studies on PT in the $Z_3$ symmetric scalar Higgs sector. Even though only four parameters $(\ld_s, \ld_{sh}, A_s, \mu_s^2)$  are 
introduced, it is still a very time-consuming job to employ a full parameter space scanning. Instead, we shall 
here focus on the possible PT scenarios indicated by the analytical analysis and demonstrate the typical 
behaviors of the viable parameter space.

\begin{figure}
\begin{center}
 \includegraphics[scale=0.6]{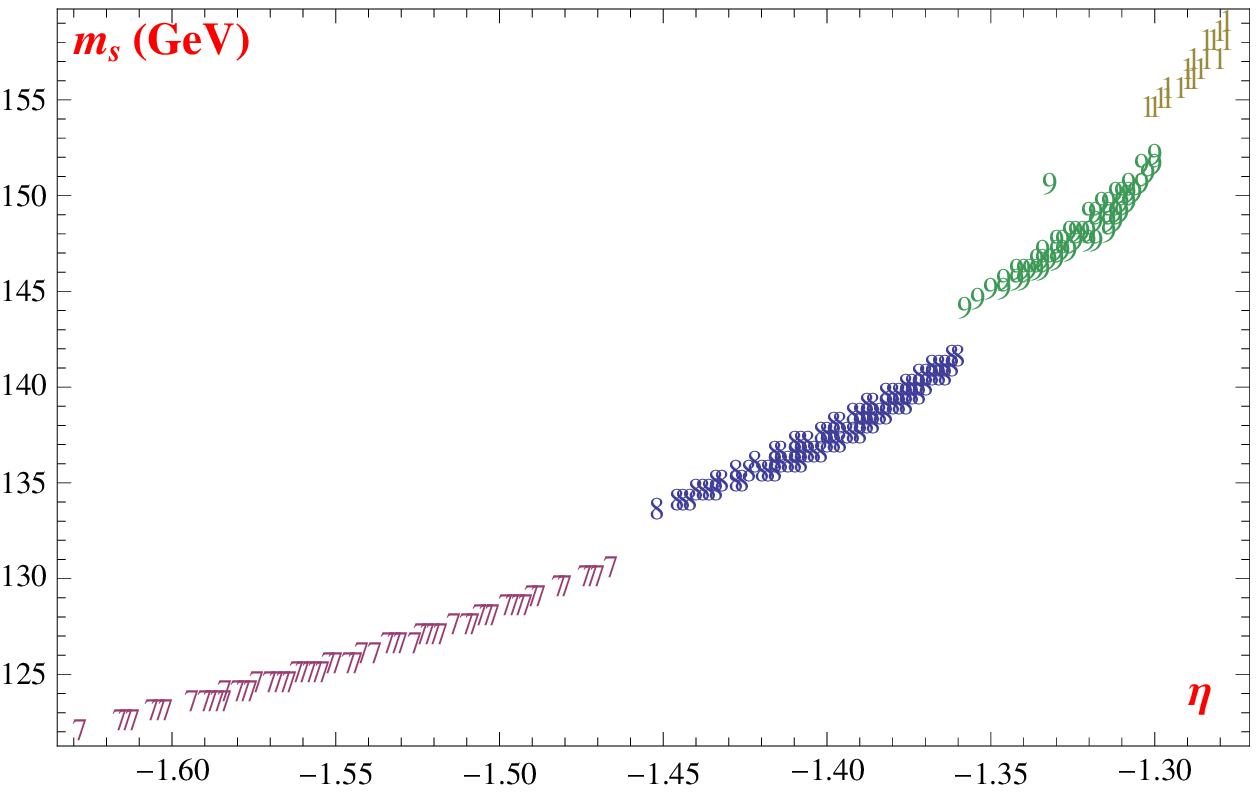}~~~
 \includegraphics[scale=0.6]{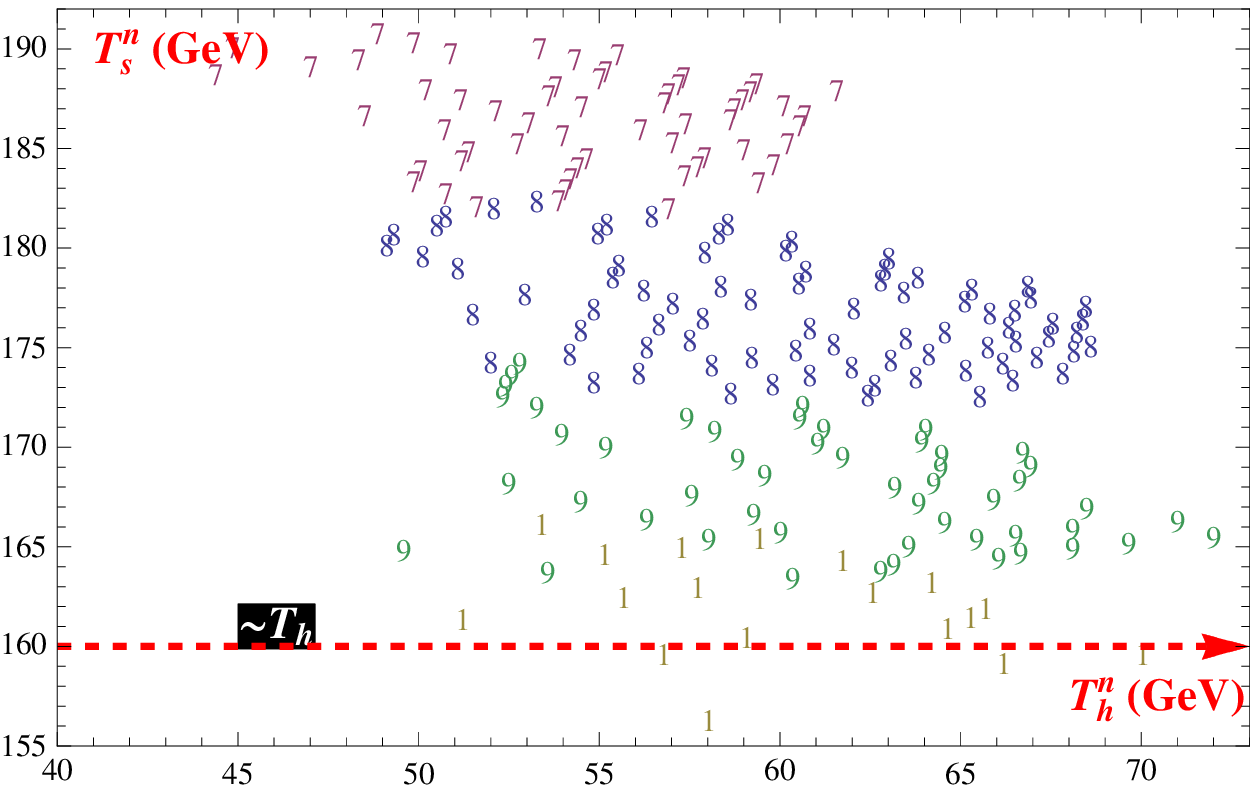} 
 \caption{\label{fig:region}
(Left panel): A demonstration on the viable three-step EWPT strips on the $(\eta , m_s)$ plane with $\ld_{sh}=0.24$, varying $\ld_s=0.7, 0.8, 0.9 ,1.0$; for easy identification we plot the region with points labelled by numbers 7, 8..., respectively. The overall picture is consistent with the analytical analysis in the high-temperature expansion. (Right panel): Distributions of two FOPT temperatures, $T_s^*$ and $T_h^*$; 
the red dashed line denotes $T_h$, the typical second order PT temperature for   
$\Omega_0\ra \Omega_h$.}  
\end{center}
\end{figure}
First we consider the $\eta<0$ case. Based on the analysis on vacuum structure, we know that it may 
allow three-step PT for a relatively large $\ld_{sh}$. But $\eta$ is supposed to lie within a narrow region 
to keep $\Omega_{sh}$ be the metalstable local minimum instead of global minimum. Whereas the condition to admit the three-step PT is even more strict. For instance, in Fig.~\ref{fig:region}, we display the feasible regions on the {$(\eta , m_s)$} plane for a given $\ld_{sh} =0.24$,~\footnote{We chosen $\ld_{sh}=0.24$. As a matter of fact, the allowed $\ld_{sh}$ cannot be significantly smaller or larger than this value. And from our preliminary searches, $\ld_{sh}\sim 0.2-0.3$.}  choosing several values of $\ld_s$; each one corresponds to a slim strip, with a shape similar to the green band shown in the left panel of Fig.~\ref{fig:vacu0}. It is clearly seen that increasing $\ld_s$ pushes $|\eta|$ towards the smaller and narrower region; at the same time, $m_s$ is in the heavier region. Eventually, for $\ld_s\simeq1.1$, the allowed region of $\eta$ is closed. Although we cannot figure out the fundamental reason for that kind of behavior, which must be a complex interplay of several effects, the direct reason is clear from the right panel of Fig.~\ref{fig:region}. Increasing $\ld_{s}$ lowers $T_s^*$ and it will eventually go below $T_h$, thus shutting down the three-step PT. On the other hand, when $\ld_{s}$ becomes fairly small (thus for a much larger $v_s$),  the regions  for $\Omega_{sh}$ in the {$ (\eta , m_s)$} plane will be occupied by that for $\Omega_s$. The reason can be traced back to the failure of the third condition in Eq.~(\ref{eq:con:2}). In addition to that, the barrier height between $\Omega_{sh}$ and $\Omega_h$ accordingly increases and thus the tunneling becomes more and more difficult. This compels $T_h^*$ to approach a fairly low temperature, an obvious trend in the left panel of Fig.~\ref{fig:vacu0}. These two effects together yield the lower bound on $\ld_{s}$, merely a little bit smaller than $0.7$ in this numerical example. 

\begin{figure}
\begin{center}
 \includegraphics[scale=0.85]{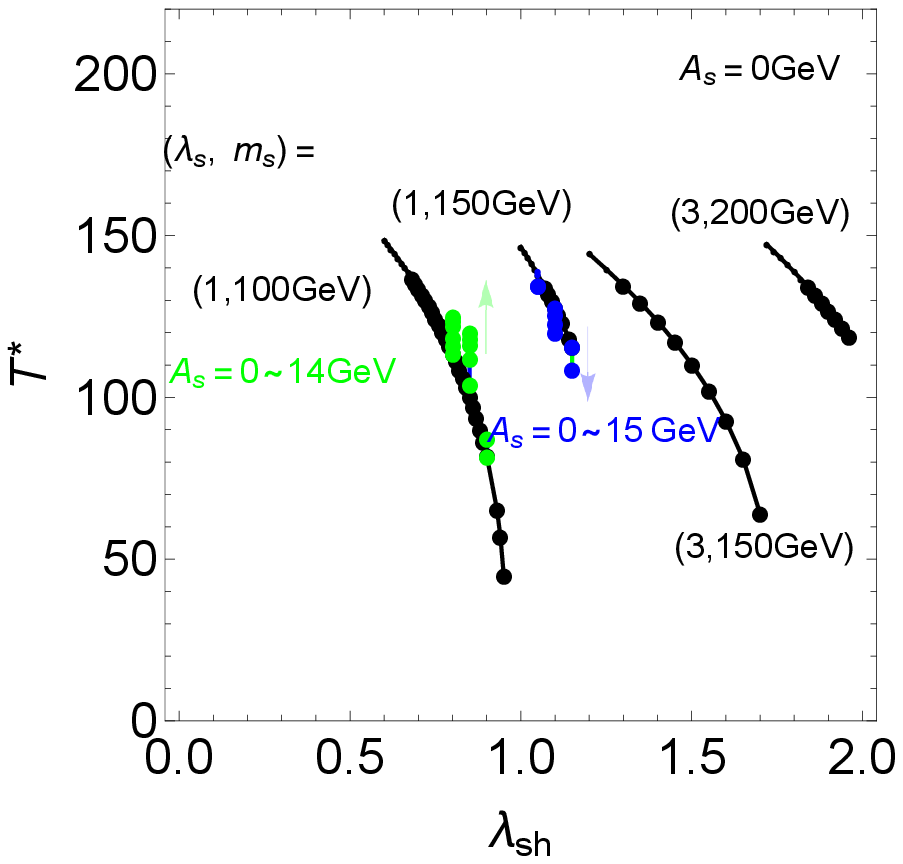}~~~
 \includegraphics[scale=0.85]{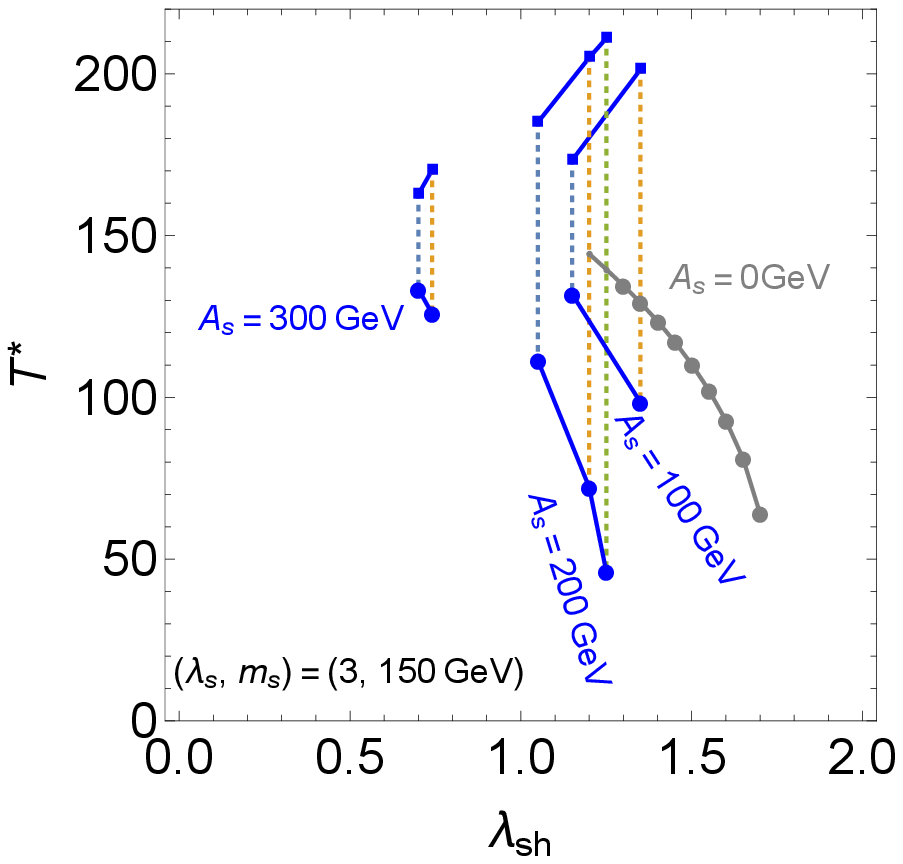}~~~
 \caption{\label{fig:lsh-T}
First order phase transition temperature as the function of $\ld_{sh}$ for the two-step PT ($\Omega_{0} \to \Omega_{s} \to \Omega_{h}$) in the $\eta>0$ region. For the  second-first order PT pattern (left panel), 
we show the $Z_2$-like case with $A_s=0$ GeV (black lines) and the deviations from non-zero $A_s$ 
by fixing $\ld_{sh}$ (green and blue lines), for four cases $(\ld_s, m_s [\GeV])=(1, 100), (1, 150), (3, 100), 
(3, 150)$. The first-first order PT pattern  (right panel) arises in the larger $\ld_s$ region which allows a 
larger $A_s$;  for $(\ld_s, m_s [\GeV])=(3, 150)$, we show $A_s [\GeV]=100, 200, 300$ (blue lines). 
For each dashed line, the upper and the lower ends denote $T_s^*$ and $T_h^*$, respectively. 
In these plots we just keep the points which give SFOEWPT.}  
\end{center}
\end{figure}
\begin{figure}
\begin{center}
\includegraphics[scale=0.85]{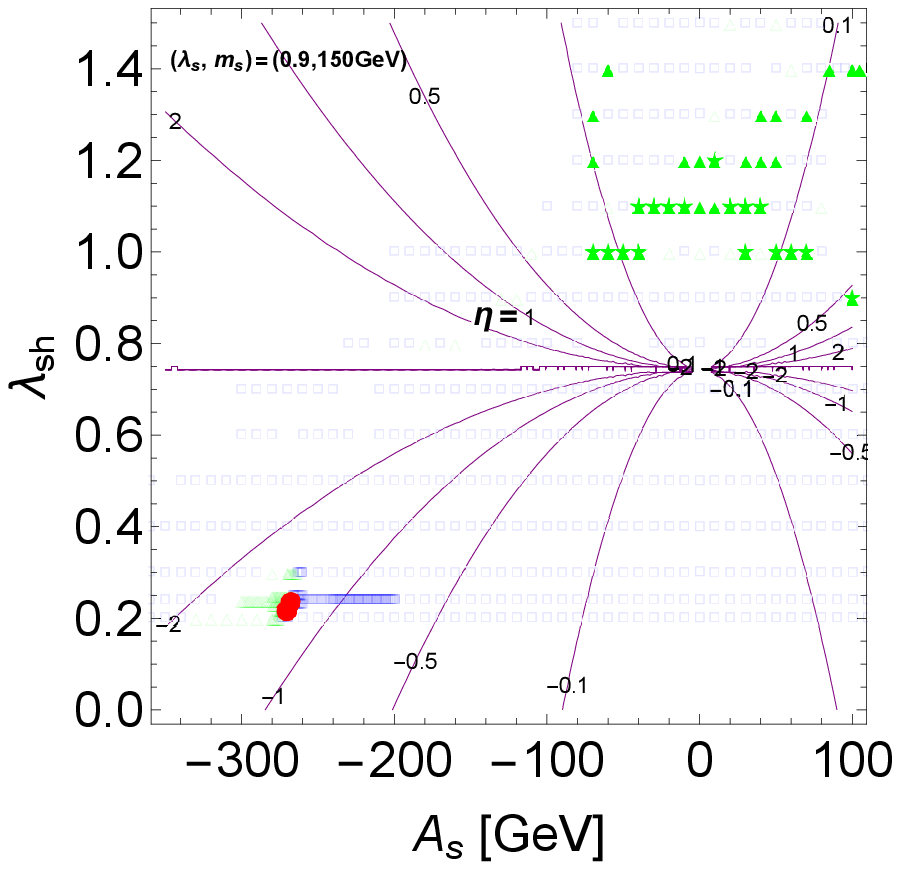}~~~
\includegraphics[scale=0.85]{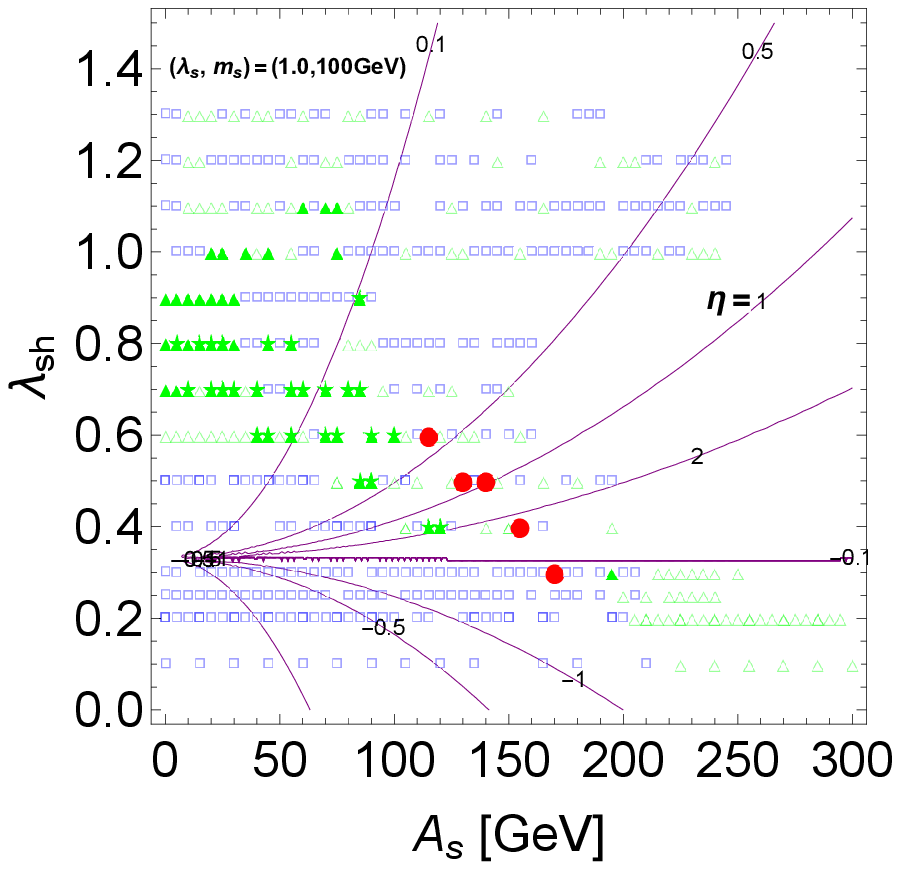}
  \caption{\label{fig:As-lsh}
Global picture of multi-step PT in the $(A_s, \ld_{sh})$ plane for $(\ld_{s}, m_s [\GeV])=(0.9, 150)$ 
(left panel) and $(1.0, 100)$ (right panel). PT of three-step (red, circle), two-step (green, triangle for 
the second-first order PT or star for the first-first order PT) and one-step (blue, square) are plotted. 
Filled plots satisfy the condition of SFOEWPT in Eq.~(\ref{eq:SFOEWPT}). 
In $\eta<0$ region, the three-step PT can happen only in a very narrow space, consistent with Fig.~\ref{fig:region}.
}
\end{center}
\end{figure}

Unlike the above case where only three-step PT is found, in the $\eta>0$ region we find that the two-step 
PT $\Omega_0\ra\Omega_s\ra\Omega_{h}$ can happen, with the first-step either second or first order, depending on the relevant parameters. We describe them one by one in the following:
\begin{description}
\item[ Second order - first oder]  This is not surprising since our model basically reduces to the $Z_2$-symmetric model in the $\eta\ra0^{+}$ limit.~\footnote{The difference is up to the expressions of $c_{h,s}$ since in our model the scalar field is complex.} First we study this limit, where the first-step is second order except for a very large $\ld_{sh}$. We obtain similar conclusions in Refs.~\cite{Curtin:2014jma,Vaskonen:2016yiu,Beniwal:2017eik,Kurup:2017dzf}: The Higgs portal coupling $\ld_{sh}$ cannot be made quite small and $\ld_s$ should also be relatively large; a large $\ld_{sh}$ is able to lower $T_h^*$ thus increasing the SOPT strength, and increasing $\ld_{s}$ could further help, shown in Fig.~\ref{fig:lsh-T}. We then investigate the role of $A_s\neq 0$ \footnote{Note that $A_s \neq 0$ tells the $Z_3$ model from the $Z_2$ model. thus being the discriminant of two models.}, which tends to make $\Omega_s$ be the global minima; see Fig.~\ref{fig:vacu1}. For the $\ld_s=1$ example, $A_s$ is restricted to be smaller than tens of GeV and thus the resulting deviations as expected are not significant. But it can still increase or decrease $T_h^*$ 
with appreciate amount, see the green and blue points in the left panel of Fig.~\ref{fig:lsh-T}. 
\item[ First order - first oder]  For a large $\ld_{s}=3$, the metastable $\Omega_s$ can be accommodated for much larger $A_s\sim{\cal O}(100)$ GeV. That large $A_s$, by contrast, is able to change the nature of transition $\Omega_0\ra\Omega_{s}$, into the first order type; furthermore, the strength of the second-step can be significantly enhanced and then reopens the smaller $\ld_{sh}$ region with $\ld_{sh} \sim$ 
(a few)$\times 0.1$; see the right panel of Fig.~\ref{fig:lsh-T}. From this figure one can find that again the requirement $T_s^*\gtrsim T_h$ yields an upper bound on $|A_s|\lesssim300$ GeV in this example. Note that the figures indicate that for a given $A_s$, the allowed region for $\ld_{sh}$ is restricted and within this region increasing $\ld_{sh}$ could again lead to a lower $T_{h}^*$. 
\item[ Three-step]  Three-step PT may be also possible from the analytical analysis and the numerical results confirm this. In fact, we find that it is much easier than the realization in the previous  scenario with $\eta<0$, where we have shown fine-tuning the parameters seems unavoidable. Whereas here it is realized in a wide parameter space, including the quite small $\ld_{sh}$ region. Since the main features are similar to the previous one, we will not add specific figures for this case. But one may gain some impression from the summarizing 
Fig.~\ref{fig:As-lsh}.
\end{description}

To have a comprehensive impression on the patterns of PT in the $Z_3$-symmetric model, we summarize the parameter region of multi-step PT in Fig.~\ref{fig:As-lsh}, where three-step PT and two-step PT are plotted. The three-step PT case for $\eta<0$ is shown in Fig.~\ref{fig:As-lsh}~(left) with $(\ld_s,  m_s~[\GeV])=(1.0, 150)$ which is corresponding to $\ld_s=0.9$ in Fig.~\ref{fig:region} with $\ld_{sh}=0.24$. 
On the other hand, we can see the region of the three-step PT case for $\eta>0$ in Fig.~\ref{fig:As-lsh}~(right). 
Most of the region of SFOEWPT is realized by the two-step PT. 
For lager $A_s$, the first-step PT significantly becomes the FOPT. 
Notice that the one-step EWPT becomes the second-order if we assume small $\ld_{sh}$. 
SFOEWPT with the one-step EWPT along the Higgs doublet direction does not appear for the parameter range in Fig.~\ref{fig:As-lsh}, 
because it is realized for $m_s \gtrsim 400$ GeV with large $\ld_{sh}$ by the non-decoupling thermal loop effects even for $A_s=0$~GeV as discussed in Refs.~\cite{Curtin:2014jma,Kakizaki:2015wua,Hashino:2016rvx,Beniwal:2017eik,Kurup:2017dzf}.

\section{Strong gravitational wave from PT with supercooling}

FOPT due to a tree level barrier typically gives rise to significant supercooling, which results in sizable free energy release during PT~\cite{Caprini:2015zlo}. Hence, strong GW is well expected in our $Z_3$ symmetric model.   As a characteristic signal in our model giving three-step PT with two FOPT, two sources of GW are furnished. However, it turns out that only the second FOPT, namely the SOEWPT is detectable in the near future. Actually, this PT pattern is totally within the  sensitivities of upcoming GW experiments such as eLIAS, DECIGO and BBO. 

\subsection{$\alpha$ \& $\beta$ description on GW from SOPT}

The GW spectrum from SOPT is very complicated, depending on quite a few details of the bubble dynamics, which is another very complicated job depending on the details of SOPT. But the complications can be parameterized by several parameters, with the most crucial two, $\alpha$ and $\beta$, which capture 
the main features of SOPT dynamics and largely determine the features of GW spectrum. We will follow the 
conventions in Ref.~\cite{Caprini:2015zlo}. 

The parameter $\alpha$ is the total energy budget of SOPT normalized by the radiative energy 
\begin{eqnarray} 
\alpha\equiv \f{\epsilon}{\rho_{\rm rad}},\quad \rho_{\rm rad}=\f{\pi^2}{30} g_* T_*^4,
\end{eqnarray}
with $g_*(=108.75)$ being  the relativistic degrees of freedom in the plasma at the PT temperature $T_*$. 
The liberated latent heat $\epsilon=-(\Delta V+T\partial V/\partial T)|_{T_*}$, with $\Delta V$ the vacuum 
energy gap between two vacua. For SOPT which typically has a significant supercooling, the latent heat 
actually is the vacuum energy. For most SFOEWPT parameter region, we have $\alpha\ll1$ owing to the 
smallness of $\Delta V$. 

Another parameter $\beta$ is defined to be the variation of action with respect to time at $T_*$: 
\begin{eqnarray} 
\beta \equiv -\left. \f{dS_3}{dt}\right\vert_{t_*}=  H_* T_* \left. \f{dS_3}{dT} \right|_{T_*},
\end{eqnarray}
with the Hubble constant $H_*\equiv 1.66 \sqrt{g_*} \, T_*^2/m_{\rm pl}$. So, its inverse characterizes the duration of PT ($\tau\sim1/\beta$) thus the GW peak frequency; usually it is much shorter than the Hubble 
time scale $1/H_*$. It is convenient to introduce the dimensionless parameter $ \tilde{\beta} \equiv \beta/
H_*\gg1$.  

The stochastic GW background in the linear approximation receive three contributions (In principle, in the case with two SOPT there are two GW sources thus six contributions.): 
\begin{eqnarray}\label{eq:GW:relic}
\Omega_{\rm GW} h^2\approx \L\Omega_{\col} h^2+\Omega_{\sw} h^2+\Omega_{\turb} h^2\R,
\end{eqnarray}
where three terms stand for relics originating from bubble collision, sound waves and 
magnetohydrodynamics (MHD) turbulence in the plasma, respectively. 
Their structures can be factorized as
\begin{eqnarray} 
\Omega_{a} h^2=c_a \L\f{1}{\wt\beta}\R^{n_a}\L\f{100}{g_*}\R^{\f{1}{3}}\L\f{\kappa_a \alpha}{1+\alpha}
\R^{r_a}f_a(v_b)S_a(f) ,
\end{eqnarray}
with $a$ denoting the subscripts in Eq.~(\ref{eq:GW:relic}). Concretely, 
\begin{itemize}
\item Bubble collision: In the envelope approximation, $c_{\col}=1.67\times10^{-5}$, $n_{\col}=2$, $r_{\col}=2$ and $f_{\col}(v_b)=0.11v_b^3/(0.42+v_b^2)$~\cite{col} with $v_b$ being the velocity of the bubble wall. 
$\kappa_{\col}$ is the fraction of latent heat deposited in a thin shell close to the PT front. 
The shape factor is defined as 
\begin{eqnarray} 
S_{\col}(f)= \f{3.8(f/f_{\col})^{2.8}}{1+2.8(f/f_{\col})^{2.8}},\quad f_{\col}=h_* \f{0.62}{1.8-0.1v_b+v_b^2}\wt \beta,
\end{eqnarray}
where $f_{\col}$ is the red-shifted peak frequency, with $h_*= 1.65\times10^{-2}{\rm mHz}\f{T_*}{100~\GeV} \L\f{g_*}{100}\R^{\f{1}{6}}$ the value of the inverse Hubble time at $T_*$ redshifted to today. An analytical treatment  to this source was made in Ref.~\cite{GW:ana}.
\item Sound waves \& MHD turbulence: Both are due to bulk motion, giving $c=2.65\times10^{-6} 
[3.35\times 10^{-4}]$, $n=1[1]$, $r=2[3/2]$ and $f(v_b)=v_b[v_b]$, respectively~\cite{sw,MHD}. 
The enhancement $H_*/\beta$ is traced back to the longer lasting time in producing GWs than the 
previous case. $\kappa_{\sw}\approx \alpha/0.73$ (for $v_b\simeq1$) is the fraction of latent heat transferred to the bulk motion of the fluid, while $\kappa_{\turb}\approx \epsilon_{\turb} \kappa_{\sw}$ with $\epsilon_{\turb}\sim5-10\%$ the fraction of bulk motion that is turbulent~\cite{sw}. The shape factors for these two cases are 
\begin{eqnarray}\label{eq:SW:P}
S_{\sw}=\L\f{f}{f_{\sw}}\R^3\L\f{7}{4+3(f/f_{\sw})^2}\R^{7/2},~~
S_{\turb}=\f{\L{f}/{f_{\turb}}\R^3}{[1+(f/f_{\turb})]^{11/3}(1+8\pi f/h_*)},
\end{eqnarray}
with the corresponding redshifted peak frequencies given by $f_{\sw[\turb]}=1.14 [1.64]  \wt \beta (h_*/v_b)$. \end{itemize}
These values obtained from simulations suffer from uncertainties and can only be fully trusted in certain 
regions of $(\alpha, v_b)$~\cite{Caprini:2015zlo}; for instance, the expression $\Omega_{\sw}h^2$ is safely 
reliable only for $\alpha\lesssim 0.1$~\cite{sw}, but we may still use it for the larger $\alpha$. 
 
\subsection{Excellent prospects of $Z_3$-symmetric model at GW detectors}

\begin{figure}
 \begin{center}
\includegraphics[scale=0.85]{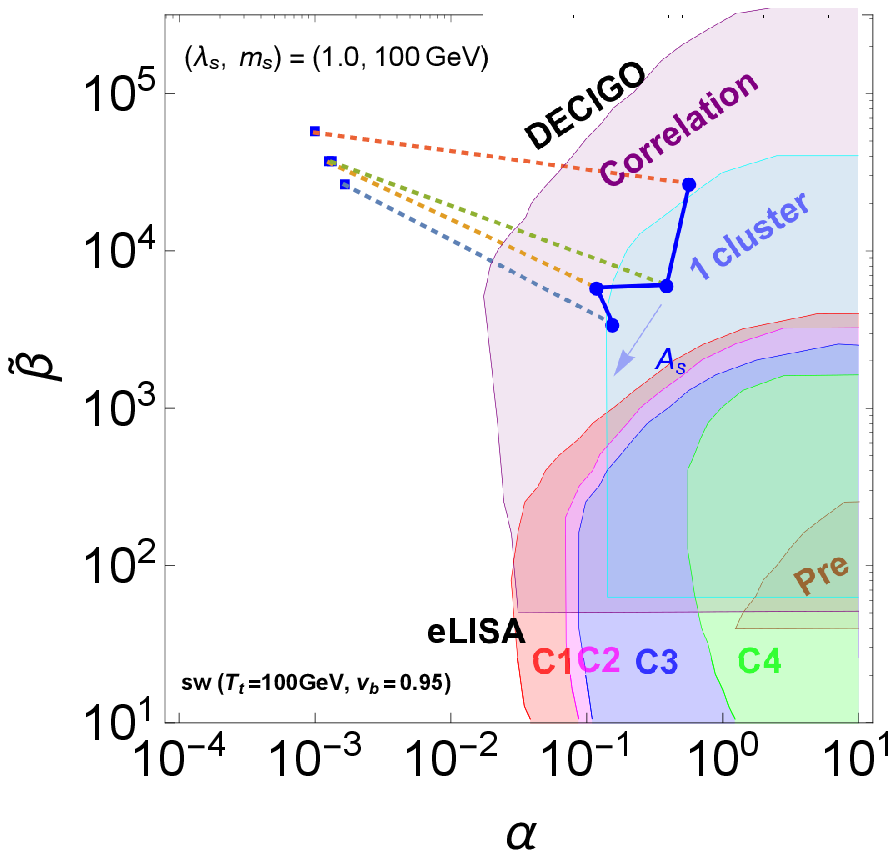}~~~
\includegraphics[scale=0.85]{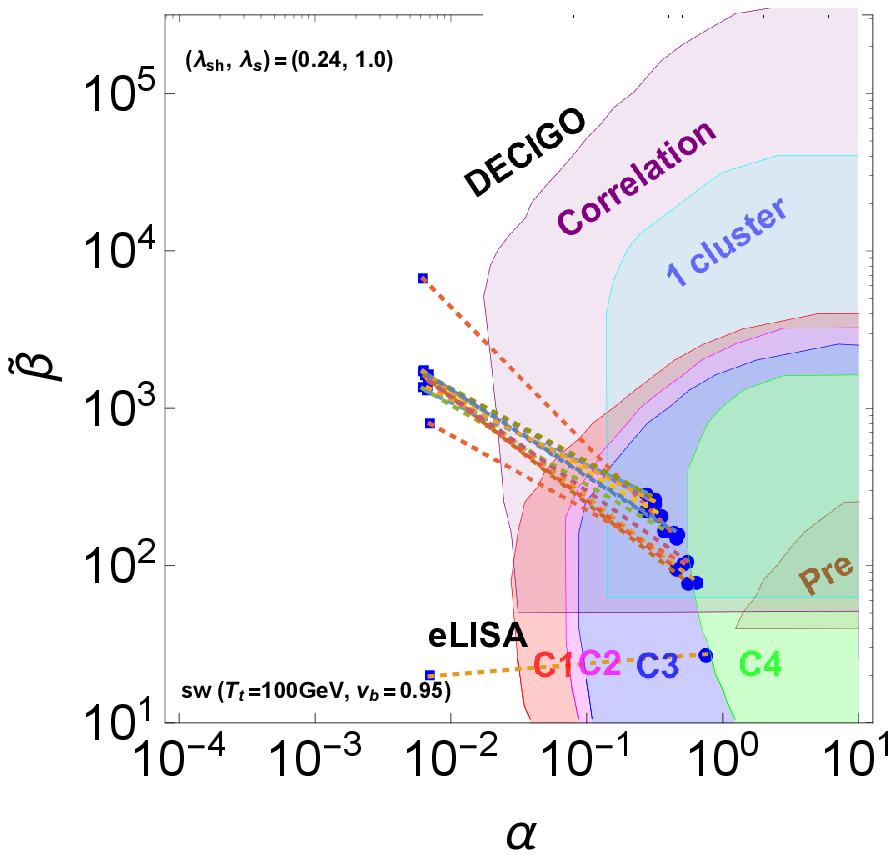}
  \caption{\label{fig:a-b_3step}
Detectability of twin GW sources from the three-step (first-second-first order) PT pattern ($\Omega_{0} \to \Omega_{s} \to \Omega_{sh} \to \Omega_{h}$) in the ($\alpha$, $\widetilde{\beta}$) plane. 
For each point, the two sources from $\Omega_{0} \to \Omega_{s}$ and $\Omega_{sh} \to \Omega_{h}$ 
are  labelled respectively by the square and circle points, connected by a dashed line. 
(Left panel for $\eta>0$): Scanning points of $\ld_{sh}$ and $\eta$ are plotted for the example 
$(\ld_{s}, m_{s} [\GeV])=(1.0, 100)$. (Right panel for $\eta<0$): Scanning points of $m_s$ and $\eta$ are plotted for the example $(\ld_{sh}, \ld_{s})=(0.24, 1.0)$. The expected experimental sensitivities of eLISA 
and DECIGO are set by using the sound wave contribution for $T_*=100$ GeV and $v_b=0.95$. 
}
\end{center}
\end{figure}
As one of the main original motivation to study the $Z_3$-symmetric singlet scalar extension, we expected that there would be two sources of GW and therefore a distinguishable twin-peak GW could show up. Qualitatively it is true and we indeed find that two SOPTs can come from both three and two-step PTs, thus contributing to two GW sources. Unfortunately, quantitatively it is disappointing. 

\begin{figure}
 \begin{center}
\includegraphics[scale=0.85]{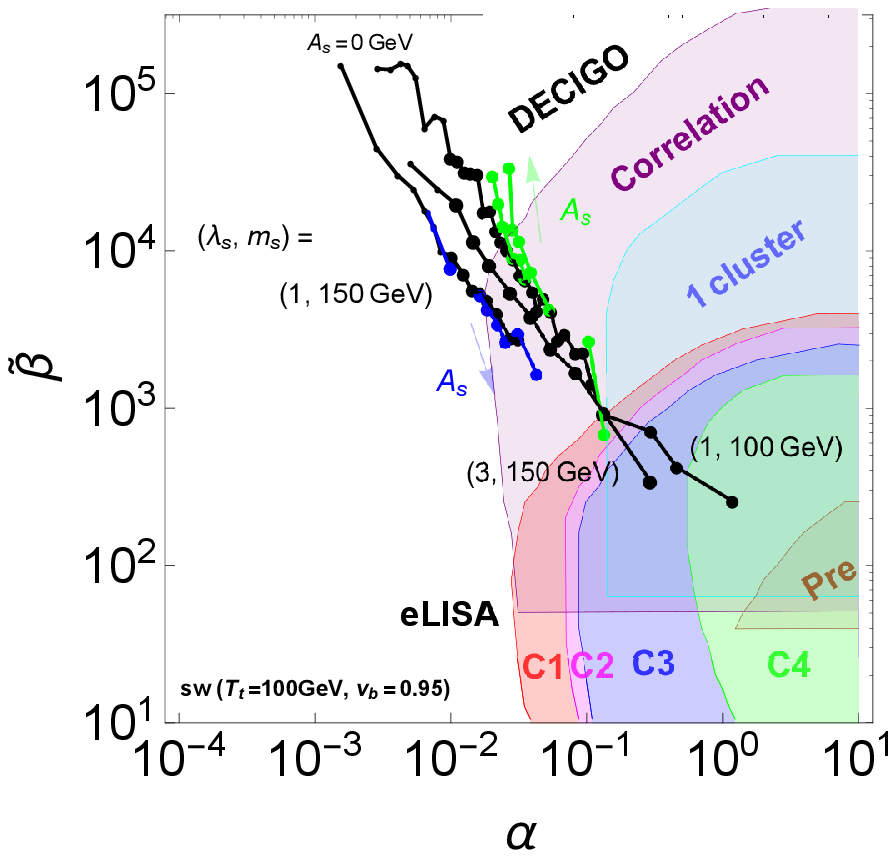}~~~
\includegraphics[scale=0.85]{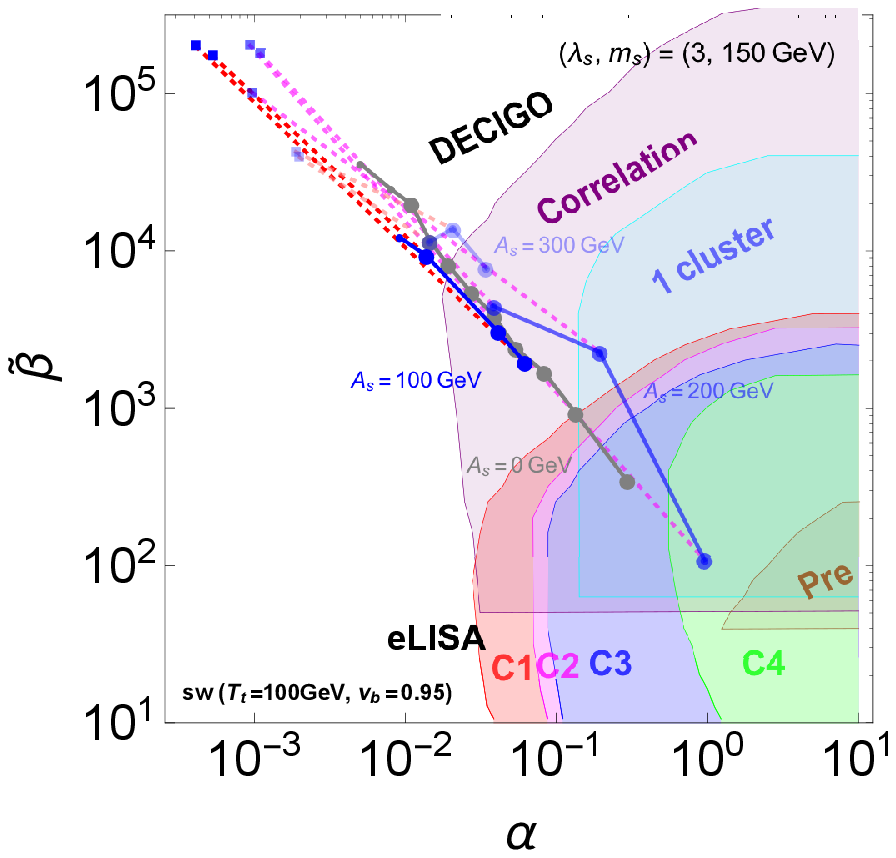}
  \caption{\label{fig:a-b_2step}
  Detectability of GWs as a function of $\ld_{sh}$ from the two-step PT ($\Omega_{0} \to \Omega_{s} \to \Omega_{h}$) in the ($\alpha$, $\widetilde{\beta}$) plan. Left panel: Second-first order pattern; Right panel: first-first order pattern. They correspond to Fig.~\ref{fig:lsh-T}.}
\end{center}
\end{figure}

We display the results on the $(\alpha, \wt\beta)$ plane in the Fig.~\ref{fig:a-b_3step} and Fig.~\ref{fig:a-b_2step}, with the experimental sensitivities of eLISA~\cite{PetiteauDataSheet,Caprini:2015zlo} and DECIGO~\cite{Kawamura:2011zz} labelled by the shaded regions. The sensitivity regions of four eLISA detector configurations described  in Table I in Ref.~\cite{Caprini:2015zlo} are denoted by ``C1'', ``C2'',
 ``C3'' and ``C4''.
The expected sensitivities for the future DECIGO stages are labeled by ``Correlation'', ``1 cluster'' and ``Pre'' following Ref.~\cite{Kawamura:2011zz}. The transition temperature $T_*^{}$ depends on the model parameters (see, Fig.~\ref{fig:region} and Fig.~\ref{fig:lsh-T}) and the velocity of the bubble wall $v_b^{}$ is uncertain. Although the experimental sensitivities on the $(\alpha, \wt\beta)$ depend on $T_*$ and $v_b^{}$, we take $T_*^{}=50~\GeV$ and $v_b=0.95$ as a reference for the purpose of illustration. It is seen that typically one needs $\alpha\gtrsim {\cal O}(0.01)$ for the near future detection. However, the first source from $\Omega_0\ra\Omega_s$ turns out to be undetectable since it always gives $\alpha\lesssim 0.01$. On the other hand, the other source, in particular in the three-step PT case, is very promising and almost the whole parameter space can be covered. One of the main reasons causing this difference is that the first-step happened at a relatively high temperature $T_s^*\gtrsim 160$ GeV, which typically is rather higher than the EWPT temperature $T_h^*\lesssim 100$ GeV; recalling that $\alpha\propto 1/T^4$, thus the first source is suppressed. A lower $T_h^*$ also leads to smaller $\wt \beta$, which is determined by the PT temperature. 



\subsection{Concerning EWBG}

SFOEWPT is a necessary but not the sufficient condition for EWBG. Despite of providing fairly SFOEWPT in the $Z_3$-symmetric limit, EWBG is not as promising as the GW detection prospect at LISA. 

The successful EWBG has a close relation with the dynamics of bubble wall. For example, the wall velocity plays key roles both to GW and EWBG, but in an opposite way. The former needs $v_b\sim 1$ to enhance the GW signals, and for a low $v_b$ all the sources are suppressed by $v_b^p$ with $p\gtrsim 3$. Whereas the latter strongly favors a lower wall velocity $v_b\lesssim 0.15-0.3$ (See the calculation of $v_b$ in the singlet models~\cite{Kozaczuk:2015owa}.), which allows the effective diffuse of particle asymmetries near the bubble wall front~\cite{diffuse}. However, a recent paper Ref.~\cite{No:2011fi} pointed out that in EWBG the relevant velocity actually is the relative velocity  ($v_+$) between the bubble wall and the plasma just in front the wall, which may be made much smaller than $v_b\lesssim c_s(=1/\sqrt{3})$ in the deflagration region with $\alpha\gtrsim{\cal O}(0.1)$. In this mechanism, the produced GW signal is detectable at eLISA and DECIGO. In the same year Ref.~\cite{Caprini:2011uz} came up with the supersonic EWBG mechanism that also operates for bubble expansion in the manner of donation with $v_b>c_s$.~\footnote{It heavily relies on the donation wave which can heat the plasma just behind the wall, raising its temperature locally above $T_c$ thus causing symmetry restoration; consequently, symmetric bubbles, where the sphaleron process violating baryon number is reactive, can be factories of baryon number.} Whereas the runaway bubble, which expands with speed of light due to the insufficient friction from the plasma, can not accommodate EWBG. It is shown that runaway bubble is likely to happen for $\alpha_*\sim{\cal O}(1)$~\cite{Espinosa:2010hh}. Therefore, from the distributions of $\alpha$ in Fig.~\ref{fig:a-b_3step} and Fig.~\ref{fig:a-b_2step} it is seen that the bubbles typically are in the deflagration scenario, but very strong SFOEWPT may result in runaway bubbles.

Regardless of the wall velocity issue, there is another serious problem which probably renders the three-step PT scenario irrelevant to EWBG. During EWPT, the phase outside the bubble, $\Omega_{sh}$, already breaks the EW symmetry and therefore the baryogensis process, which should be effective outside the bubble, actually was ineffective due to the suppression on sphaleron rate by $e^{-E_{sph}(T_h^*)/T_h^*}\sim e^{-\f{4\pi}{g_2} \langle h_*\rangle/T_h^*}\ll1$. However, we cannot completely excludes this case, since the idea of locally recovering EW symmetry used in the supersonic EWBG might work here, and it deserves a specific study elsewhere.







\section{Conclusions and discussions}

Gravitational waves, which are supposed to be relics of strong first order PT, is a good probe to new physics beyond SM complementary to collider searches. Within SM, EWPT is second order, while EWBG requires SFOEWPT. Such a situation inspires a great many extensions to the SM Higgs sector where a potential barrier is created during EWPT by, e.g., non-thermal tree level effects. One simple implement of this idea is in the mixing Higgs-singlet models which includes a sizable cubic term like $S|H|^2$~\cite{Profumo:2007wc,Noble:2007kk,Ashoorioon:2009nf,2step1,Fuyuto:2014yia,Profumo:2014opa,Chen:2014ask,Kanemura:2015fra,Tenkanen:2016idg,Kanemura:2016lkz,Huang:2016cjm,Hashino:2016xoj}, and then a doublet-singlet mixing could contribute to enhance the strength of the FOPT. As a result, such kinds of models can be tested by the synergy between the measurements of various Higgs boson couplings at future collider experiments and the observation of GWs at future space-based interferometers as discussed in Refs.~\cite{Huang:2016cjm,Hashino:2016xoj}. In another implementation imposing unbroken discrete symmetry like $Z_2$~\cite{2step1,2step2,Curtin:2014jma,Vaskonen:2016yiu,Chao:2017vrq,Beniwal:2017eik,Kurup:2017dzf}, multi-step PT could utilize a tree level barrier. But generically the absence of mixing renders the tests at colliders very difficult without taking enough large $\ld_{sh}$ coupling as discussed in Refs.~\cite{Curtin:2014jma,Kakizaki:2015wua,Hashino:2016rvx,Beniwal:2017eik,Kurup:2017dzf}. In this paper, we have focused on such the nightmare scenario. 

Following the second line, in this paper we  studied in great detail the patterns of SFOEWPT in the next to simplest extension to the SM Higgs sector, the $Z_3$-symmetric extension to the Higgs sector. It involves one more term, the cubic term for the singlet (as a comparison, the $Z_2$ model can be a limit of turning off the corresponding coupling $A_s$), that gives rise to remarkable difference in PT than the $Z_2$-symmetric model. It can not only significantly enhance the potential barrier thus the PT strength, but also lead to three-step PT through the intermediate phase $\Omega_{sh}$ in contrast with the $Z_2$-symmetric model. SFOEWPT is expected to be fairly strong due to the significant supercooling. Especially, the three-step PT produces two sources of GW. Despite of the undetectability from the first-step in the near future, the other source with significant supercooling and thus leads to $\alpha\sim{\cal O}(0.1)$, basically can be completely covered by eLISA and DECIGO. Additionally, there were also studies on FOPT thus GWs associated with the singlet scalar, without taking into account EWPT~\cite{EGW:1,EGW:2}. 

We end up the paper with some open questions. First, we encounter some new phenomenologies/problems in multi-step SFOEWPT, which needs further understanding. Second, if the three-step PT in our model can give sizable baryon number asymmetry is unclear and should be studied elsewhere, although it would depend on the specific models for baryogenesis involving a new source of CP violation. In the last, the SOPT phenomenologies are quite rich in this model, and if we give up the $Z_3$-symmetric ground state, there are even more intriguing scenarios. For instance, the pattern $\Omega_0\ra\Omega_h\ra\Omega_{sh}$ is in the bulk parameter and has important consequence to dark matter. We leave this part to future publication.


\section{Acknowledgements}

This work is supported in part by National Research Foundation of Korea (NRF) Research Grant NRF-2015R1A2A1A05001869 (PK, TM).

\appendix




\vspace{-.3cm}
\begin{thebibliography}{99}


\bibitem{EWBG} 
V. A. Kuzmin, V. A. Rubakov, and M. E. Shaposhnikov, 
Phys. Lett. B155 (1985) 36; M. E. Shaposhnikov, 
Nucl. Phys. B287 (1987) 757-775.

\bibitem{Morrissey:2012db} 
  D.~E.~Morrissey and M.~J.~Ramsey-Musolf,
  New J.\ Phys.\  {\bf 14}, 125003 (2012).

\bibitem{Chung:2012vg} 
  D.~J.~H.~Chung, A.~J.~Long and L.~T.~Wang,
  Phys.\ Rev.\ D {\bf 87}, no. 2, 023509 (2013).

\bibitem{Patel:2011th} 
  H.~H.~Patel and M.~J.~Ramsey-Musolf,
  JHEP {\bf 1107}, 029 (2011).



\bibitem{NMSSM:PT1} 
M. Pietroni, Nucl. Phys. B402, 27 (1993). 

\bibitem{NMSSM:PT2} 
  R.~Apreda, M.~Maggiore, A.~Nicolis and A.~Riotto,
  Nucl.\ Phys.\ B {\bf 631}, 342 (2002). 

\bibitem{NMSSM:PT3} 
A. Menon, D. E. Morrissey, and C. E. M. Wagner, Phys.
Rev. D 70, 035005 (2004).

\bibitem{NMSSM:PT4} 
K. Funakubo, S. Tao, and F. Toyoda, Prog. Theor. Phys.
114, 369 (2005).

\bibitem{Profumo:2007wc} 
  S.~Profumo, M.~J.~Ramsey-Musolf and G.~Shaughnessy,
  JHEP {\bf 0708}, 010 (2007).

\bibitem{Noble:2007kk} 
A. Noble and M. Perelstein, Phys. Rev. D78, 063518 (2008). 

\bibitem{Ashoorioon:2009nf} 
  A.~Ashoorioon and T.~Konstandin,
  JHEP {\bf 0907}, 086 (2009).

\bibitem{2step1} 
  J.~R.~Espinosa, T.~Konstandin and F.~Riva,
  Nucl.\ Phys.\ B {\bf 854}, 592 (2012).

\bibitem{NMSSM:PT5} 
M. Carena, N. R. Shah, and C. E. M. Wagner, Phys. Rev. D
85, 036003 (2012).

\bibitem{Huang:2014ifa} 
  W.~Huang, Z.~Kang, J.~Shu, P.~Wu and J.~M.~Yang,
  Phys.\ Rev.\ D {\bf 91}, no. 2, 025006 (2015).

\bibitem{multiPT} 
H. H. Patel and M. J. Ramsey-Musolf, Phys. Rev. D88 (2013) 035013; 
C. Cheung and Y. Zhang, JHEP 09 (2013) 002; 
  N.~Blinov, J.~Kozaczuk, D.~E.~Morrissey and C.~Tamarit,
  Phys.\ Rev.\ D {\bf 92}, no. 3, 035012 (2015).


\bibitem{Fuyuto:2014yia} 
  K.~Fuyuto and E.~Senaha,
  Phys.\ Rev.\ D {\bf 90}, no. 1, 015015 (2014). 

\bibitem{Profumo:2014opa} 
  S.~Profumo, M.~J.~Ramsey-Musolf, C.~L.~Wainwright and P.~Winslow,
  Phys.\ Rev.\ D {\bf 91}, no. 3, 035018 (2015).

\bibitem{Chen:2014ask} 
  C.~Y.~Chen, S.~Dawson and I.~M.~Lewis,
  Phys.\ Rev.\ D {\bf 91}, no. 3, 035015 (2015).





\bibitem{Kanemura:2015fra} 
  S.~Kanemura, M.~Kikuchi and K.~Yagyu,
  Nucl.\ Phys.\ B {\bf 907}, 286 (2016)

\bibitem{Tenkanen:2016idg} 
  T.~Tenkanen, K.~Tuominen and V.~Vaskonen,
  JCAP {\bf 1609}, no. 09, 037 (2016).

\bibitem{Kanemura:2016lkz} 
  S.~Kanemura, M.~Kikuchi and K.~Yagyu,
  Nucl.\ Phys.\ B {\bf 917}, 154 (2017).

\bibitem{Huang:2016cjm} 
  P.~Huang, A.~J.~Long and L.~T.~Wang,
  Phys.\ Rev.\ D {\bf 94}, no. 7, 075008 (2016).

\bibitem{Hashino:2016xoj} 
  K.~Hashino, M.~Kakizaki, S.~Kanemura, P.~Ko and T.~Matsui,
  Phys.\ Lett.\ B {\bf 766}, 49 (2017).

\bibitem{Bian:2017wfv} 
  L.~Bian, H.~K.~Guo and J.~Shu,
  arXiv:1704.02488 [hep-ph].



\bibitem{2step2} 
J. M. Cline and K. Kainulainen, 
JCAP 1301 (2013) 012.
   
 \bibitem{Curtin:2014jma} 
  D.~Curtin, P.~Meade and C.~T.~Yu,
  JHEP {\bf 1411}, 127 (2014).

\bibitem{Chala:2016ykx}
M.~Chala, G.~Nardini and I.~Sobolev,
with direct detection and gravitational wave signatures,''
Phys.\ Rev.\ D {\bf 94}, no. 5, 055006 (2016).

\bibitem{Vaskonen:2016yiu} 
  V.~Vaskonen,
  Phys.\ Rev.\ D {\bf 95}, no. 12, 123515 (2017).

\bibitem{Chao:2017vrq} 
  W.~Chao, H.~K.~Guo and J.~Shu,
  arXiv:1702.02698 [hep-ph].

\bibitem{Beniwal:2017eik} 
  A.~Beniwal, M.~Lewicki, J.~D.~Wells, M.~White and A.~G.~Williams,
  arXiv:1702.06124 [hep-ph].




\bibitem{Land:1992sm} 
  D.~Land and E.~D.~Carlson,
  Phys.\ Lett.\ B {\bf 292}, 107 (1992). 

\bibitem{Hammerschmitt:1994fn} 
  A.~Hammerschmitt, J.~Kripfganz and M.~G.~Schmidt,
  Z.\ Phys.\ C {\bf 64}, 105 (1994).

 \bibitem{Kurup:2017dzf} 
  G.~Kurup and M.~Perelstein,
  arXiv:1704.03381 [hep-ph].

\bibitem{Chao:2017oux} 
  W.~Chao,
  arXiv:1706.01041 [hep-ph].



\bibitem{Kanemura:2004ch} 
  S.~Kanemura, Y.~Okada and E.~Senaha,
  Phys.\ Lett.\ B {\bf 606}, 361 (2005).

\bibitem{AKS}
  M.~Aoki, S.~Kanemura and O.~Seto,
  Phys.\ Rev.\ Lett.\  {\bf 102}, 051805 (2009);
  Phys.\ Rev.\ D {\bf 80}, 033007 (2009);
  M.~Aoki, S.~Kanemura and K.~Yagyu,
  Phys.\ Rev.\ D {\bf 83}, 075016 (2011).

\bibitem{Kanemura:2011fy} 
  S.~Kanemura, E.~Senaha and T.~Shindou,
  Phys.\ Lett.\ B {\bf 706}, 40 (2011). 

\bibitem{Kobakhidze:2015xlz} 
  A.~Kobakhidze, L.~Wu and J.~Yue,
  JHEP {\bf 1604}, 011 (2016).

\bibitem{Tamarit:2014dua} 
  C.~Tamarit,
  Phys.\ Rev.\ D {\bf 90}, no. 5, 055024 (2014). 

\bibitem{Kanemura:2014cka} 
  S.~Kanemura, N.~Machida and T.~Shindou,
  Phys.\ Lett.\ B {\bf 738}, 178 (2014). 

\bibitem{Hashino:2015nxa} 
  K.~Hashino, S.~Kanemura and Y.~Orikasa,
  Phys.\ Lett.\ B {\bf 752}, 217 (2016). 

\bibitem{Kakizaki:2015wua} 
  M.~Kakizaki, S.~Kanemura and T.~Matsui,
  Phys.\ Rev.\ D {\bf 92}, no. 11, 115007 (2015). 

\bibitem{Hashino:2016rvx} 
  K.~Hashino, M.~Kakizaki, S.~Kanemura and T.~Matsui,
  Phys.\ Rev.\ D {\bf 94}, no. 1, 015005 (2016).

\bibitem{Lewis:2017dme}
I.~M.~Lewis and M.~Sullivan,
  arXiv:1701.08774 [hep-ph].


\bibitem{Chen:2017qcz}
C.~Y.~Chen, J.~Kozaczuk and I.~M.~Lewis,
  arXiv:1704.05844 [hep-ph].




\bibitem{ILC}
%
  J.~Brau, (Ed.) {\it et al.}  [ILC Collaboration],
  arXiv:0712.1950 [physics.acc-ph];
%
  G.~Aarons {\it et al.}  [ILC Collaboration],
  arXiv:0709.1893 [hep-ph];
%
  N.~Phinney, N.~Toge and N.~Walker,
  arXiv:0712.2361 [physics.acc-ph];
%
  T.~Behnke, (Ed.) {\it et al.}  [ILC Collaboration],
  arXiv:0712.2356 [physics.ins-det];
%
T.~Behnke {\it et al.},
arXiv:1306.6329 [physics.ins-det];
%
H.~Baer, {\it et al.} "Physics at the International Linear Collider", 
{\it Physics Chapter of the ILC Detailed Baseline Design Report}:
http://lcsim.org/papers/DBDPhysics.pdf.

\bibitem{CLIC}
  E.~Accomando {\it et al.}  [CLIC Physics Working Group Collaboration],
  hep-ph/0412251;
  L.~Linssen, A.~Miyamoto, M.~Stanitzki and H.~Weerts,
  arXiv:1202.5940 [physics.ins-det].

\bibitem{ILCHiggsWhitePaper} 
  D.~M.~Asner, T.~Barklow, C.~Calancha, K.~Fujii, N.~Graf, H.~E.~Haber, A.~Ishikawa and S.~Kanemura {\it et al.},
  arXiv:1310.0763 [hep-ph].

\bibitem{Moortgat-Picka:2015yla} 
  G.~Moortgat-Pick {\it et al.},
  Eur.\ Phys.\ J.\ C {\bf 75}, no. 8, 371 (2015).

\bibitem{Fujii:2015jha} 
  K.~Fujii {\it et al.},
  arXiv:1506.05992 [hep-ex].


\bibitem{GWs:early1} 
A. Kosowsky, M. S. Turner, and R. Watkins, Phys. Rev. Lett. 69, 2026 (1992); Phys. Rev. D 45, 4514 (1992). 

\bibitem{GWs:early2} 
M. Kamionkowski, A. Kosowsky, and M. S. Turner, Phys. Rev. D 49, 2837 (1994).

\bibitem{GWs:early3} 
A. Kosowsky, A. Mack, and T. Kahniashvili, Phys. Rev. D
66, 024030 (2002); A. D. Dolgov, D. Grasso, and A. Nicolis, Phys. Rev. D 66,
103505 (2002).



\bibitem{Caprini:2015zlo} 
  C.~Caprini {\it et al.},
  JCAP {\bf 1604}, no. 04, 001 (2016).

\bibitem{Cai:2017cbj} 
  R.~G.~Cai, Z.~Cao, Z.~K.~Guo, S.~J.~Wang and T.~Yang,
  doi:10.1093/nsr/nwx029
  arXiv:1703.00187 [gr-qc]. 



\bibitem{aLIGO}
  B.~P.~Abbott {\it et al.} [LIGO Scientific and Virgo Collaborations],
  Phys.\ Rev.\ Lett.\  {\bf 116}, no. 6, 061102 (2016);
  Phys.\ Rev.\ Lett.\  {\bf 116}, no. 24, 241103 (2016); 
%
  B.~P.~Abbott {\it et al.} [LIGO Scientific and VIRGO Collaborations],
  Phys.\ Rev.\ Lett.\  {\bf 118}, no. 22, 221101 (2017). 

\bibitem{Harry:2010zz} 
  G.~M.~Harry [LIGO Scientific Collaboration],
  Class.\ Quant.\ Grav.\  {\bf 27}, 084006 (2010).

\bibitem{Seoane:2013qna} 
  P.~A.~Seoane {\it et al.} [eLISA Collaboration],
  arXiv:1305.5720 [astro-ph.CO].

\bibitem{Kawamura:2011zz} 
  S.~Kawamura {\it et al.},
  Class.\ Quant.\ Grav.\  {\bf 28}, 094011 (2011).
  
\bibitem{Corbin:2005ny} 
  V.~Corbin and N.~J.~Cornish,
  Class.\ Quant.\ Grav.\  {\bf 23}, 2435 (2006).



\bibitem{Ma:2007gq} 
  E.~Ma,
  Phys.\ Lett.\ B {\bf 662}, 49 (2008).

\bibitem{Kang:2010ha} 
  Z.~Kang and T.~Li,
  JHEP {\bf 1102}, 035 (2011).

\bibitem{Belanger:2012zr} 
  G.~Belanger, K.~Kannike, A.~Pukhov and M.~Raidal,
  JCAP {\bf 1301}, 022 (2013).

\bibitem{Ko:2014nha} 
  P.~Ko and Y.~Tang,
  JCAP {\bf 1405}, 047 (2014);
  P.~Ko and Y.~Tang,
  JCAP {\bf 1501}, 023 (2015).
  

\bibitem{Aoki:2014cja} 
  M.~Aoki and T.~Toma,
  JCAP {\bf 1409}, 016 (2014).

\bibitem{Guo:2015lxa} 
  J.~Guo, Z.~Kang, P.~Ko and Y.~Orikasa,
  Phys.\ Rev.\ D {\bf 91}, no. 11, 115017 (2015).

\bibitem{Cai:2016hne} 
  Y.~Cai and A.~Spray,
  JHEP {\bf 1702}, 120 (2017)

\bibitem{Baker:2016xzo} 
  M.~J.~Baker and J.~Kopp,
  arXiv:1608.07578 [hep-ph].


\bibitem{Arcadi:2017vis} 
  G.~Arcadi, F.~S.~Queiroz and C.~Siqueira,
  arXiv:1706.02336 [hep-ph].

\bibitem{Ding:2016wbd} 
  R.~Ding, Z.~L.~Han, Y.~Liao and W.~P.~Xie,
  JHEP {\bf 1605}, 030 (2016)


  \bibitem{Wainwright:2011kj} 
  C.~L.~Wainwright,
  Comput.\ Phys.\ Commun.\  {\bf 183}, 2006 (2012).


 \bibitem{GW:ana} 
R.Jinno and M.Takimoto, PRD95(2017) no.2, 024009. 



\bibitem{BubbleN} 
M. S. Turner, E. J. Weinberg and L. M. Widrow, Phys. Rev. D 46 (1992) 2384.

\bibitem{Espinosa:2010hh} 
  J.~R.~Espinosa, T.~Konstandin, J.~M.~No and G.~Servant,
  JCAP {\bf 1006}, 028 (2010).

\bibitem{col} 
S. J. Huber and T. Konstandin, JCAP 0809 (2008) 022. 


Ryusuke

\bibitem{MHD} 
C. Caprini, R. Durrer and G. Servant, JCAP 0912 (2009) 024; 
P. Binetruy, A. Bohe, C. Caprini and J. F. Dufaux, JCAP 1206 (2012) 027. 

\bibitem{sw} 
M. Hindmarsh, S. J. Huber, K. Rummukainen and D. J. Weir, Phys. Rev. D 92 (2015)
12, 123009. 

\bibitem{PetiteauDataSheet}
Data sheet by A. Petiteau, \\
http://www.apc.univ-paris7.fr/Downloads/lisa/eLISA/Sensitivity/Cfgv1/StochBkgd/

\bibitem{diffuse} 
M. Joyce, T. Prokopec and N. Turok, Phys. Rev. Lett. 75 (1995) 1695 [Erratum-ibid. 75 (1995) 3375]; Phys. Rev. D 53 (1996) 2930; Phys. Rev. D 53 (1996) 2958.

 \bibitem{No:2011fi} 
  J.~M.~No,
  Phys.\ Rev.\ D {\bf 84}, 124025 (2011).

\bibitem{Kozaczuk:2015owa} 
  J.~Kozaczuk,
  JHEP {\bf 1510}, 135 (2015). 


\bibitem{Caprini:2011uz} 
  C.~Caprini and J.~M.~No,
  JCAP {\bf 1201}, 031 (2012).


\bibitem{EGW:1} 
"Gravitational waves from the first order phase transition of the Higgs field at high energy scales"
R.Jinno, K.Nakayama and M.Takimoto, Phys. Rev. D 93 (2016) no.4, 045024. 

\bibitem{EGW:2} 
  L.~Marzola, A.~Racioppi and V.~Vaskonen,
  arXiv:1704.01034 [hep-ph].


 \end{thebibliography}
\end{document}